\pdfoutput=1
\documentclass[conference,compsoc]{IEEEtran}

\usepackage{tabularx}
\usepackage{booktabs}
\usepackage{url}
\usepackage{subcaption}
\usepackage{fontawesome5}

\usepackage{tikz}
\usepackage[T1]{fontenc}
\usepackage{amsmath}
\usetikzlibrary{babel}
\usepackage{pifont}
\usepackage{listings}
\usepackage{xspace}
\usepackage{url}
\usepackage{ifthen}
\usepackage{xcolor}
\usepackage{array}
\usepackage{listings}
\usepackage{colortbl}
\usepackage{adjustbox}
\usepackage[utf8]{inputenc}
\usepackage{twemojis}
\usepackage{multirow}
\usepackage[most]{tcolorbox}

\usepackage{float}
\usepackage{placeins}

\usetikzlibrary{arrows.meta}

\makeatletter
\g@addto@macro{\UrlBreaks}{\do\A\do\B\do\C\do\D\do\E\do\F\do\G\do\H\do\I\do\J\do\K\do\L\do\M\do\N\do\O\do\P\do\Q\do\R\do\S\do\T\do\U\do\V\do\W\do\X\do\Y\do\Z\do\a\do\b\do\c\do\d\do\e\do\f\do\g\do\h\do\i\do\j\do\k\do\l\do\m\do\n\do\o\do\p\do\q\do\r\do\s\do\t\do\u\do\v\do\w\do\x\do\y\do\z\do\0\do\1\do\2\do\3\do\4\do\5\do\6\do\7\do\8\do\9\do\-}
\makeatother

\newcommand{\para}[1]{\smallskip \noindent \textbf{#1}}

\usepackage{makecell}
\usepackage{cleveref}
\usepackage{graphicx}
\crefformat{section}{\S#2#1#3}
\crefformat{subsection}{\S#2#1#3}
\crefformat{subsubsection}{\S#2#1#3}

\newcommand{\eg}{e.g.,\xspace}
\newcommand{\ie}{i.e.,\xspace}

\newcommand{\tool}{\textsc{SST-Guard}\xspace}

\newcommand{\totalDetected}{6,314\xspace}

\definecolor{navyblue}{RGB}{0, 0, 128}

\newcommand*\blackcircled[1]{%
  \tikz[baseline=(char.base)]{
    \node[
      shape=circle,
      fill=black,
      text=white,
      font=\scriptsize\bfseries,
      minimum size=1.1em,
      text width=1.1em,
      align=center,
      inner sep=0pt
    ] (char) {#1};
  }%
}

\definecolor{cookiecolor}{RGB}{205, 133, 63}
\definecolor{windowcolor}{RGB}{70, 130, 180}
\definecolor{networkcolor}{RGB}{60, 179, 113}

\newcolumntype{N}{>{\raggedright\arraybackslash}p{3.5cm}}
\newcolumntype{R}{>{\ttfamily\footnotesize\raggedright\arraybackslash}p{9.5cm}}

\definecolor{okblue}{HTML}{0072B2}
\definecolor{okorange}{HTML}{E69F00}
\definecolor{okpurple}{HTML}{CC79A7}
\definecolor{okgreen}{HTML}{009E73}
\definecolor{okgray}{HTML}{555555}

\newcounter{takeaway}
\newtcolorbox{takeaway}{
  enhanced, colback=gray!6, colframe=black, boxrule=0pt, leftrule=2pt,
  sharp corners, left=5pt, right=4pt, top=3pt, bottom=3pt,
  fontupper=\small, before upper={\textbf{Takeaway~\stepcounter{takeaway}\thetakeaway:}~}}

\newcommand{\bs}{\textbackslash}

\newcommand{\cmark}{\ding{51}}

\begin{document}

\title{\tool : Detecting and Characterizing Server-Side Google Analytics in the Wild}

\author{\IEEEauthorblockN{Muhammad Jazlan, Alexander Gamero-Garrido, Zubair Shafiq, Yash Vekaria}
\IEEEauthorblockA{University of California, Davis}}

\maketitle

\begin{abstract}
As web browsers increasingly implement anti-tracking features, the web tracking ecosystem has been shifting from the client-side to the server-side.
Instead of sending requests directly to the tracker's endpoint, server-side tracking (SST) sends tracking requests to publisher-controlled or intermediary endpoints that then forward the information to trackers from the server-side.
As a result, client-side tracking protections are becoming fragile because direct client-to-tracker requests may no longer be observed.

In this paper, we investigate the server-side implementation of Google Analytics (sGA), the most widely deployed third-party tracking service on the web today.
We present \tool, a multi-modal browser-based approach for detecting sGA.
The key insight behind \tool is that common sGA deployments change the standard Google Analytics endpoints, but still leave semantic artifacts of data collection by Google Analytics in the browser, including identifiers, event metadata, cookies, and JavaScript state.
\tool detects these artifacts using value-templates, which use regular expressions to match semantic value patterns across multiple modalities: network requests, cookies, and the \texttt{window} object, providing robustness against obfuscation of one modality.

We validate \tool on Tranco top-10k websites, detecting 4.0\% (401) sGA domains with over 93.0\% accuracy across three modalities, with network request classifier demonstrating the highest accuracy (99.8\%).
Additionally, \tool identifies atleast 43.2\% more websites that integrate sGA as compared to prior approaches.
Deploying \tool at scale, we detect sGA on 4.2\% (\totalDetected) of Tranco top-150K websites.
Our analysis shows that many sGA deployments use first-party subdomains, direct A/AAAA records, custom paths, or encoded payloads that circumvent existing defenses.
We reported our findings to the authors of EasyPrivacy, who have since integrated them.

\end{abstract}

\begin{IEEEkeywords}
Web Privacy, Server Side Tracking, Google Analytics
\end{IEEEkeywords}

\section{Introduction}
\label{sec:introduction}

Online tracking typically operates through client-side requests to third-party endpoints of advertising and analytics services, allowing trackers to observe, aggregate, and correlate user activity across distinct first-party websites (\ie publishers)~\cite{vekaria2025sok}.
Approximately 90\% of websites embed at least one third party, with Google and Meta among the most prevalent third parties, present on 53\% and 22\% of all websites~\cite{WebAlmanac.2025.ThirdParties}.
The inclusion of these third parties poses a privacy risk for users whose personal information may be shared with advertising and analytics services \cite{flo2021FTC, betterhelp2023FTC, wilson2024DMV}.

To defend against third-party tracking, major browsers have introduced tracking protection features targeting \textit{client-side} tracking~\cite{SafariPrivacyPage, mozillaEnhancedTracking, BraveLists, BraveShields}.
Other tracking protection tools use filter lists (\eg EasyList, EasyPrivacy) to block tracking requests in the browser~\cite{easylist,easyprivacy}.
Together, these interventions have led to a growing ``signal loss'' for advertising and tracking services~\cite{iab2024revenue}.
In response, trackers and publishers have increasingly moved tracking into the first-party context, using techniques such as first-party cookies~\cite{chen2021cookie, demir2022towards, munir2023cookiegraph, nikkhah2025cookieguard} and CNAME cloaking~\cite{dao2021cname, dimova2021cname}.
For example, Dimova et al.~\cite{dimova2021cname} found CNAME cloaking on 9.98\% of the top 10K websites, with its use growing by 21\% over 22 months.

More recently, a new tracking paradigm called \textit{server-side tracking} (SST) has emerged as a more flexible form of first-party tracking integration~\cite{fouad2024devil, elfraihi2024serverside, vekaria2025sok}.
Traditionally, a third-party tracking script embedded on a website directly reports data from the user's browser to the tracker's endpoint.
SST decouples data collection from data reporting: while tracking data is still collected in the user's browser, the browser sends the resulting tracking requests to a publisher-controlled or an intermediary endpoint rather than directly to the third-party tracker.
The intermediary then forwards the tracking payloads to the actual tracking service through server-to-server communication.
As a result, SST removes the direct client-to-tracker request that client-side tracking protections rely on to detect and block tracking.
SST has now been adopted and promoted by major advertising and analytics platforms, including Google~\cite{google_sgtm_why_when}, Meta~\cite{meta_capi_about}, and others~\cite{snap_capi, reddit_capi, tiktok_events_api}.

SST undermines user privacy by disguising data collection as first-party traffic, preventing users from identifying embedded trackers or blocking them with conventional tools.
Furthermore, because SST cannot rely on third-party cookies, it shifts cross-site tracking toward mechanisms users cannot easily control, such as first-party cookies, hashed email addresses, and browser fingerprinting.
Ultimately, while SST offers publishers server-side data governance, it yields negligible privacy benefits over existing client-side configurations at the significant cost of user transparency and control.
Thus, accurate detection of SST is essential to protect users from this emergent form of tracking.

However, detecting SST is challenging for four reasons.
First, publishers can route tracking data through arbitrary first-party endpoints, making it hard to block them without breaking the website's functionality.
Second, these tracking requests can be customized to use publisher-specific paths and non-standard parameter names.
Third, requests sent to intermediaries can be encoded or obfuscated to prevent simple rule-based blocking.
Fourth, there is no reliable ground truth for SST, since tracking data is forwarded server-side, outside the browser.

Related work on detecting SST relies on signals that publishers can change, which limits the ability of anti-tracking tools built on them.
Fouad et al.~\cite{fouad2024devil} conducted the first dedicated study of SST, detecting trackers that migrate from client-side to server-side by comparing crawls from before and after the introduction of server-side Google Tag Manager (sGTM), which requires a historical crawl and misses path-based deployments.
Moti et al.~\cite{moti2025bitter} identify tracking requests by matching known parameter names, which fails once a publisher renames, removes, or obfuscates them.
Concurrent to our work, Mertens et al.~\cite{mertens2026detecting} detect Collector Tags configured with an sGTM URL in GTM containers, which misses deployments that do not use sGTM.

Our work focuses on detecting server-side Google Analytics (sGA), the most prevalent third party on the web.
In contrast to prior work, we posit that sGA deployments still leave browser-visible artifacts of tracking, regardless of which destination they are sent to.
When a publisher deploys sGA, the endpoint and parameter names may change, but the implementation still needs to collect and transmit the same information as client-side Google Analytics, such as identifiers and event metadata.
The \textit{values} of these artifacts retain their format even when their names are changed.
Based on this insight, we propose \tool, a novel approach of detecting sGA.
\tool uses \textit{value templates}, \ie regular expressions that match the values of Google Analytics artifacts rather than their names.
\tool applies value templates across three modalities: network requests, cookies, and JavaScript \texttt{window} variables.

We bootstrap the training process for \tool on Tranco top-10K websites using labels obtained from client-side Google Analytics, and evaluate it on 403 sGA domains validated using Google's Tag Assistant extension~\cite{googleGoogleAssistant}.
Despite never being trained on sGA requests, \tool detects sGA requests with 98.25\% precision and 97.52\% recall.
We then compare \tool with prior and concurrent work on a fresh crawl containing 479 sGA websites (\Cref{tab:intro-comparison}).
\tool detects the exact sGA endpoint on 95.0\% of these websites, showing a 43.2 percentage point increase in detection.
When considering all three modalities, \tool detects sGA on 99.4\% of these websites.
We then deploy \tool in-the-wild on Tranco top-150K websites and detect sGA on \totalDetected websites.
81.6\% of these websites route sGA through a distinct subdomain, while the remaining 18.4\% use path-based endpoints on the first-party site.
Among subdomain-based deployments, only 26.5\% use CNAME cloaking, with the remaining 73.5\% relying on direct A/AAAA records.
82.7\% of these A/AAAA subdomains are hosted on infrastructure different from the publisher's main website.
We also find that publishers customize their sGA deployments.
28.1\% of websites use non-standard path endpoints, such as \url{/ag/g/c}, and 25.6\% heavily customize their sGA requests.
Publishers also rename the window variables set by Google scripts.
We observe 260 different names for \texttt{dataLayer}, the default variable for Google Analytics.
Finally, 223 websites base64-encode the query parameters of their sGA requests.
\tool detects 216 (96.9\%) of these websites through cookies and \texttt{window} variables, even though the request classifier alone detects only 7.

Existing defenses provide incomplete protection against sGA.
EasyPrivacy~\cite{easyprivacy} blocks 93.50\% of sGA requests, but fails to block any sGA request on 166 websites.
We reported our findings to the maintainers of EasyPrivacy, who have since integrated the most common sGA subdomain names and paths into the list.
In summary, we make the following contributions:
\begin{itemize}
    \item We propose \tool, a novel approach of detecting server-side Google Analytics using value templates across network requests, cookies, and \texttt{window} variables, without relying on known endpoints or parameter names.
    \item We show that \tool outperforms prior approaches by 43.2 percentage points for detecting validated sGA endpoints.
    \item We conduct the largest crawl for detecting sGA to date, on Tranco top-150K websites, characterizing how publishers customize their implementations.
    \item We report our findings to the maintainers of EasyPrivacy, resulting in the addition of 26 rules that are now used by millions of users.
\end{itemize}

\begin{table}[b]
\centering
\caption{Comparison with prior and concurrent work on detecting sGA.}
\label{tab:intro-comparison}
\footnotesize
\setlength{\tabcolsep}{0pt}
\begin{tabular*}{\columnwidth}{@{\extracolsep{\fill}}lcccc@{}}
\toprule
 & \makecell{Fouad\\et al.~\cite{fouad2024devil}} & \makecell{Moti\\et al.~\cite{moti2025bitter}} & \makecell{Mertens\\et al.~\cite{mertens2026detecting}} & \tool \\
\midrule
sGTM            & \cmark & \cmark & \cmark & \cmark \\
Non-sGTM        &        &        &        & \cmark \\
Renamed params  &        &        &        & \cmark \\
Path-based      &        &        &        & \cmark \\
\bottomrule
\end{tabular*}
\end{table}

\section{Background}
\label{sec:background}

\subsection{Tracking Shift Towards the Server-side}
Tracking has been a fundamental component of online advertising that allows advertisers to perform analytics, personalization, re-targeting, and conversion optimization \cite{vekaria2025sok}.
Publishers often embed third-party trackers on their websites, allowing them to execute code, maintain user state via browser storage mechanisms (\eg cookies), and report data to tracking servers directly from the \textit{client-side}.

Over the past decade, major browsers have integrated privacy enhancing tools within their browsers \cite{SafariPrivacyPage, mozillaEnhancedTracking, BraveShields}, restricting third-party tracking by default.
At the same time, privacy conscious users have relied on tools like adblockers that use community maintained filterlists to identify and block known client-side trackers \cite{AboutAdBlockPlus,uBlockOrigin,easylist,easyprivacy}.
These protections are effective: for instance, a recent work found that removing third-party cookies from a browser reduced the tracker's ability to re-identify users by over 50\% \cite{elfraihi2024serverside}.

\begin{figure}[t]
    \centering
    \includegraphics[width=\linewidth]{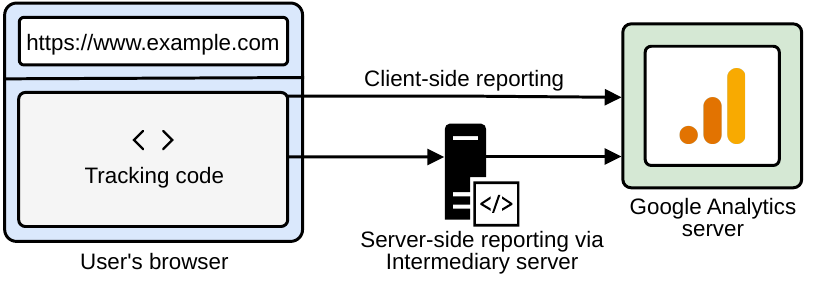}
    \caption{In client-side tracking (top), the browser sends tracking requests directly to a third party (GA server in our case).
    In server-side tracking (bottom), the browser sends tracking requests to an intermediary server, which subsequently forwards the tracking data to the GA server.}
    \label{fig:ga-client-side-architecture}
\end{figure}

However, this signal loss has prompted responses from ad-tech stakeholders.
In a recent blog post, the Interactive Advertising Bureau (IAB) wrote, ``Given the privacy related changes made and proposed by top browsers, it is clear that client side ad technology dependent on third party code execution on the client side cannot be sustained as it stands today.''
The IAB further notes that existing privacy measures in browsers ``steadily diminish user agent surface area and third party signals, severely impacting traditional ad monetization models and yields,'' and advocates for a transition to server-side architectures that ``eliminate client-side dependencies and mitigate ad-blocking threats'' \cite{iabtechlabTechTrusted}.

\begin{figure*}[ht]
    \centering
    \includegraphics[width=\linewidth]{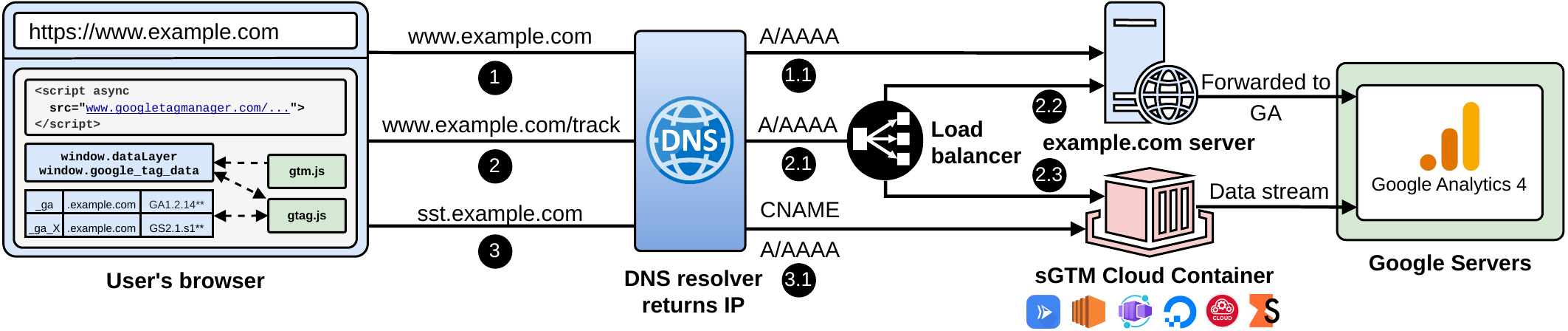}
    \caption{Typical deployment scenarios for server-side tracking via Google Analytics.
    \protect\blackcircled{1} \texttt{gtag.js} sends events to a publisher-specified first-party endpoint.
    \protect\blackcircled{2} Requests are routed through a load balancer (\protect\blackcircled{2.1}) to either a first-party server (\protect\blackcircled{2.2}) or a virtual sGTM container on the publisher's infrastructure (\protect\blackcircled{2.3}).
    \protect\blackcircled{3} The publisher deploys the sGTM container on a cloud provider, pointing a first-party subdomain via \texttt{CNAME} or \texttt{A/AAAA} records.
    In all scenarios, data is forwarded server-side to Google Analytics.}
    \label{fig:server-side-implementations}
\end{figure*}

Server-side tracking (SST) has thus emerged as a server-based architecture that addresses these concerns.
In SST, instead of the browser communicating directly with a third-party endpoint, the publisher routes tracking payloads through an intermediary server (e.g., \textit{\url{example.com/track}} instead of \textit{\url{google-analytics.com/track}}).
The intermediary server then processes and forwards these requests to the actual third-party destination via server-to-server communication.
This process is shown in Figure \ref{fig:ga-client-side-architecture}.

When the intermediary is configured on a first-party domain, browser-based protections as well as community-maintained filter lists are rendered ineffective as they primarily target third-party tracking \cite{SafariPrivacyPage, BraveShields}.
Historically, blocking a widespread third-party tracker required only a single rule (\eg \texttt{||google-analytics.com}), which protected users across every site that embedded it.
With SST, however, each publisher can integrate the tracking intermediary under a different domain, subdomain or path.
In the worst case, this fragmentation forces filter lists to maintain a separate blocking rule per website, rendering static, domain-based blocking non-scalable and largely ineffective.
We discuss further implications to privacy in \Cref{subsec:discussion:privacy-concerns}.

\subsubsection{Industry Adoption}
\label{subsubsec:background:industry-adoption}
In August 2020, Google became the first major advertiser to introduce server-side Google Analytics, an SST solution implemented via server-side Google Tag Manager \cite{simoahavaServersideTaggingLaunchBlog}, claiming to operate in a ``first-party context'' and enabling cookies to be ``more durable'' \cite{google_sgtm_why_when}.
Following Google, other advertising platforms have also adopted SST implementations, allowing publishers to track and target their website visitors on third-party platforms.
Meta released Conversions API (CAPI)~\cite{capiLaunchDateBlog}, stating that ``data from the Conversions API is less impacted than the Meta Pixel by browser loading errors, connectivity issues and ad blockers'' \cite{meta_capi_about}.
Microsoft advertises their Conversion API (CAPI) by stating that it performs even ``when client-side tracking is limited by browser restrictions or ad blockers''\cite{microsoftCAPI}.
TikTok also makes similar claims about their Events API, stating ``connectivity issues and browser inconsistency can impact conversions reported via Pixel. With Events API, more conversions are reported and leveraged for measurement, optimization, and targeting'' \cite{tiktok_events_api}.
Similarly, Reddit advertises its server-side connections to be ``more resilient to signal loss...'' on their Conversions API page \cite{reddit_capi} and Snapchat advocates their Conversions API to ``better inform the optimization and delivery of campaigns, driving more efficient cost per action'' \cite{snap_capi}.
More recently, Netflix and OpenAI have also offered Conversion API products \cite{netflixCAPI, openaiCAPI}.
Overall, these platforms often promote using both client- and server-side tracking simultaneously to improve ad delivery and targeting, capture missed conversions, and reduce cost per result \cite{meta_capi_about, tiktok_events_api}.

\subsubsection{Scope}
In this paper, we focus on detecting server-side Google Analytics (sGA) for three main reasons.
First, Google Analytics is the most prevalent third party on the web, present on 53\% of all websites~\cite{WebAlmanac.2025.ThirdParties}.
Second, sGA is often bundled with server-side Google Tag Manager (sGTM) which can be used to load multiple tags at the same time.
For example, one instance of sGTM can be used to implement sGA, Meta CAPI, TikTok Events API, Snapchat CAPI, and Reddit CAPI, simultaneously~\cite{googleIntroductionTagging}.
Therefore, detecting sGA may also, as a side-effect, identify the first-party endpoint through which publishers may share data with these other platforms.
Third, Google provides Tag Assistant~\cite{googleGoogleAssistant}, a debugging tool that reports the destination of GA events, giving us a reliable source of ground truth for sGA that, to the best of our knowledge, is not available for other server-side trackers.

\subsection{Tracking via Google Analytics}
\label{subsec:background:google-analytics}
Google Analytics 4 (GA4) is Google's current analytics platform.
Events can be reported to Google Analytics in three ways: via the Global Site Tag (\texttt{gtag.js}), Google Tag Manager (GTM), and the Measurement Protocol.
Since the Measurement Protocol is a server-to-server reporting solution that does not require any client-side code, and is not visible to Tag Assistant, we do not target it in this paper.
GA4 can be deployed as both a client-side and a server-side analytics solution.

\subsubsection{Initializing Google Analytics}
\label{subsubsec:loading-ga-code}
Google provides GA4 code with \texttt{gtag.js}, Google's unified site tag, fetched from Google servers.
For client-side implementations, a request is made to \textit{\url{googletagmanager.com/gtag/js}}.
For server-side implementations, Google recommends serving this file from a first-party endpoint like \textit{\url{example.com/gtag/js}}.
For websites using Google Tag Manager (or sGTM), the \texttt{gtm.js} file is downloaded first.
Either script (shown within the user's browser in \Cref{fig:server-side-implementations}) sets up state by initializing the \texttt{dataLayer} and \texttt{google\_tag\_data} variables.
It also sets two first-party cookies, \texttt{\_ga} and \texttt{\_ga\_<ID>}, where ID is a website-specific Measurement ID (\eg \texttt{G-ABCDEF1234}), assigned when GA4 is configured on the GA dashboard.
To use sGA, the publisher points the Google tag to the intermediary endpoint by setting the \texttt{server\_container\_url} or \texttt{transport\_url} configuration parameter, either in \texttt{gtag.js} directly or in the Google tag's configuration settings in GTM.

\subsubsection{Reporting Events}
\label{subsubsec:reporting-events}
Publishers can configure events through the GA Dashboard or by calling the \texttt{dataLayer.push} method using Javascript.
GA provides some pre-defined events (Standard Events) and allows publishers to add custom events.
When a tracking event fires on the page, details about the event are pushed onto the \texttt{dataLayer} variable and processed by the script.
For client-side GA, a network request is sent to \textit{\url{analytics.google.com/g/collect}} or \textit{\url{google-analytics.com/g/collect}}.
For  server-side GA, a network request is sent to a first-party subdomain or endpoint, which can then take any one of the routes shown in \Cref{fig:server-side-implementations}, before being reported to Google Analytics.

\begin{figure*}[t]
\centering
\includegraphics[width=\textwidth]{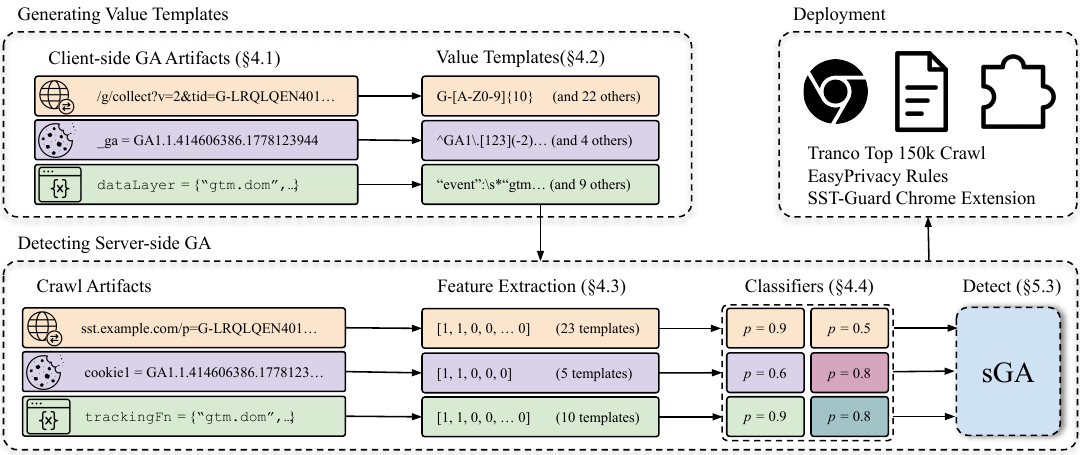}
\caption{Overview of \tool. Value templates learned from client-side GA artifacts (top) are applied to request, cookie, and window artifacts on sGA sites (bottom). Each classifier, one per modality plus combined and meta ensembles over all modalities, makes its own sGA decision.
We deploy \tool on Tranco top 150k domains, contribute rules to EasyPrivacy, and create a Chrome Extension~\cite{sstguard}.}
\label{fig:system-design}
\end{figure*}

\subsubsection{Debugging GA and sGA via Google Tag Assistant}
\label{subsubsec:background:tag-assistant}
\begin{figure}[b]
    \centering
    \includegraphics[width=0.5\linewidth]{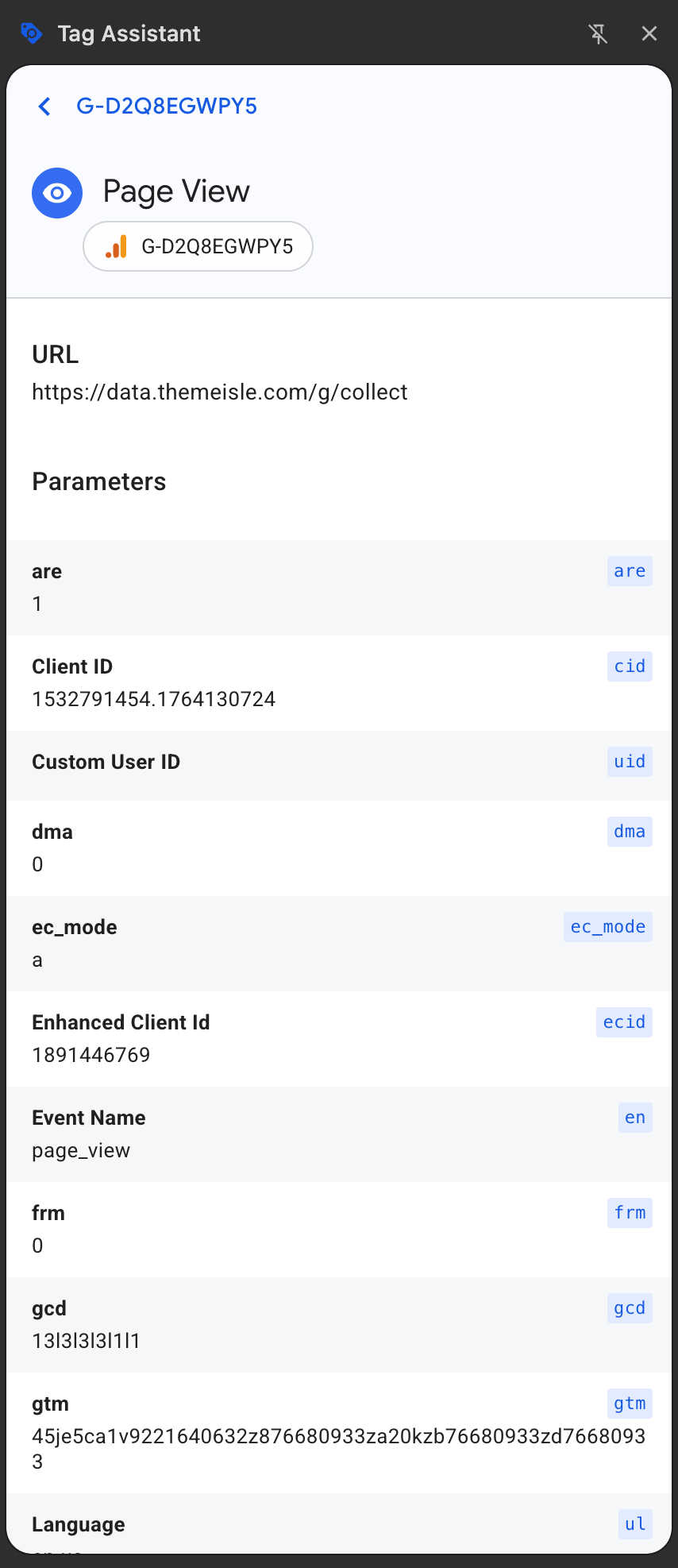}
    \caption{The Google Tag Assistant interface on a website that employs server-side tracking.}
    \label{fig:tag-assistant-screenshot}
\end{figure}

Google Tag Assistant~\cite{googleGoogleAssistant} is Google's \textit{official} browser extension that allows developers to debug their client-side as well as server-side implementations of Google Analytics and Google Tag Manager.
Whenever an event is triggered on the page, the extension fires a custom DOM event named \texttt{TAG\_ASSISTANT\_API\_MESSAGE}, which contains a \texttt{GTAG\_HIT} object with the full request URL, the destination Measurement ID, and the event name.
Since the extension captures these details directly from the Google tag, it provides reliable ground truth for both client-side and server-side GA.
\Cref{fig:tag-assistant-screenshot} shows Tag Assistant in use on \textit{\url{themeisle.com}}, a website that implements SST.
However, since Google can modify or deprecate Tag Assistant at any time, we do not rely on it to train \tool, and use it only for validation.

One limitation of Tag Assistant is that it only captures events sent through the Google tag (\texttt{gtag.js}), which all GA4 implementations use, including those deployed via GTM.
Implementations that report directly via the Measurement Protocol, such as Google Analytics 4 Server-Side offered by Freshpaint~\cite{freshpaintGoogleAnalytics}, are not captured.
Therefore, our ground truth represents a lower bound on sGA implementations.

\subsection{Flexibility in sGA Deployments}
\label{subsec:background:sga-deployment}
Client-side Google Analytics (GA) is typically deployed as a plug-and-play solution provided by Google.
However, sGA provides multiple customization and deployment options.
Publishers can override default GA behaviors (\eg request parameters) since requests are routed through publisher-controlled infrastructure before being reported to Google.
They can also choose to deploy sGA with or without sGTM, as shown in \Cref{fig:server-side-implementations}.

\para{Tracking Customization}
\label{subsubsec:sga-customization}
When using sGA, publishers can choose to arbitrarily exclude default parameters from a request.
They can further encode or transform request payloads to reduce their detectability.
For instance, Stape.io~\cite{stapeAvoidBlocking,stapeLoadContainer} offers multiple sGTM templates with this functionality.
Publishers can also rename the \texttt{dataLayer} and \texttt{google\_tag\_data} variables initialized by the Google tag.
Similarly, they can replace the default \texttt{/g/collect} endpoint with a custom path.
Finally, publishers can change cookie names, \eg via the \texttt{cookie\_prefix} configuration parameter.

\para{Deployment Infrastructure}
\label{subsubsec:background:deployment-infra}
A publisher typically deploys their website via a hosting service on a Virtual Private Server (VPS) offered by cloud providers like Google Cloud Platform, AWS, or Microsoft Azure.
A request to \textit{\url{www.example.com}} resolves to an A/AAAA record for a VPS, which returns a response to the client (step \blackcircled{1}).
Once a publisher implements sGA on their website, they need a first-party endpoint that receives tracking requests from the client.

If a publisher opts to use sGTM, they need to deploy an sGTM cloud container on a dedicated infrastructure.
To receive requests from the client-side, they also need to add a CNAME or A/AAAA record for their domain.
If a publisher opts to deploy sGA without sGTM, they do not need to deploy any additional infrastructure.
However, they do need to implement additional logic on the server-side to forward their tracking data to Google Analytics.

There are two primary ways in which this infrastructure can be added: first, the publisher uses a specific path for tracking requests (step \blackcircled{2}), or second, the publisher uses a distinct subdomain (step \blackcircled{3}).
These choices impact transparency into sGA deployments.
For instance, a distinct subdomain is more transparent since its DNS records reveal further details (\eg who owns the tracking subdomain, where it is deployed).
However, if a publisher hides their implementation behind a load balancer, transparency decreases, since we cannot determine where the tracking information is transmitted.

\section{Threat Model}
\label{sec:threat-model}
We consider an adversary comprising first-party publishers and third-party advertising and tracking platforms.
We assume that publishers deploy an SST solution offered by the tracking platform on their website to track users.
We assume that users want to remain private, and may rely on their browser's built-in privacy defenses or use anti-tracking tools such as ad blockers.

The adversary's goal is to collect analytics signals from the user's browser while making the tracking traffic indistinguishable from benign first-party traffic, thereby evading the privacy protections employed by users.
The adversary controls the SST infrastructure, including request routing, parameter naming, payload formats, and encoding schemes.
The publisher can also customize tag configurations, override default parameters, and deploy tracking endpoints on first-party domains, subdomains, or paths, using either CNAME records or direct A/AAAA records.

\tool runs in the user's browser and observes only client-side artifacts: network requests, cookies, and window variables.
We assume no access to server-side configurations, server-side logs, or any cooperation from the adversary.
Our goal is to detect sGA as it is deployed in the wild, including the customizations and obfuscation we observe in practice.

\section{The \tool Design}
\label{sec:methodology}

In this section, we explain the design of \tool, shown in \Cref{fig:system-design}.
\tool detects sGA by matching the values of GA artifacts in the browser, rather than known parameter, cookie, or variable names, and learns to do so entirely from client-side GA.
First, we explain our crawling process (\Cref{subsec:methodology:browser-instrumentation}).
Second, we introduce our \textit{Value Template} approach for matching these values (\Cref{subsec:methodology:value-templates}).
Next, we describe how we extract features (\Cref{subsec:methodology:feature-extraction}) and train our classifiers using labels bootstrapped from client-side GA (\Cref{subsec:methodology:classifier-training}).
Finally, we describe how we deploy \tool as a browser extension (\Cref{subsec:methodology:deployment}).

\subsection{Crawler Setup}
\label{subsec:methodology:browser-instrumentation}
We automate the crawling process to study server-side tracking by instrumenting Google Chrome (v138) with Playwright~\cite{playwright}, an open-source browser automation framework, on a Linux 5.15.0 machine running on Ubuntu 22.04 LTS.
For each website we crawl, we first navigate to the homepage and wait for the page to load.
We use the Consent-O-Matic~\cite{consent-o-matic} extension, configured to accept all cookies.
Our goal is to simulate real user behavior and ensure that analytics events are triggered during the visit.
To this end, once the page loads, we execute a sequence of human-like scrolls to the bottom of the page, make random pauses, and then scroll back to the top.
We also click on random non-interactive elements on the page.
Finally, we navigate to a random same-domain link on the homepage.

We continuously capture network requests, responses, request payloads, and headers using Playwright's \texttt{request} object.
We obtain a request's initiator using the Chrome DevTools Protocol (CDP).
After performing the interactions, we also collect cookies using Playwright's \texttt{storageState()} and dump all the window variables by iterating over each variable associated with the \texttt{window} object.
We use this instrumentation to crawl the Tranco list~\cite{trancoList} of top-10K domains (ID: ZW96G) and store the collected data in a structured database for further analysis.

\begin{table*}[t]
    \centering
    \small
    \caption{Combined performance metrics for Validation and Evaluation datasets. Thresholds (Thr.) represent the optimal values tuned for each classification level.}
    \begin{tabular}{l ccccc @{\hskip 0.2in} | @{\hskip 0.2in} ccccc}
        \toprule[1.1pt]
        & \multicolumn{5}{c}{\textbf{Validation Metrics}} & \multicolumn{5}{c}{\textbf{Evaluation Metrics}} \\
        \cmidrule(r){2-11}
        \textbf{Modality} & \textbf{Accuracy} & \textbf{Precision} & \textbf{Recall} & \textbf{F1} & \textbf{Thr.} & \textbf{Accuracy} & \textbf{Precision} & \textbf{Recall} & \textbf{F1} & \textbf{Thr.} \\
        \midrule
        Network Request & & & & & & & & & & \\
        \hspace{1em} Request-level & 99.72 & 73.36 & 89.24 & 80.52 & 0.9 & \textbf{99.80} & 98.25 & 97.52 & \textbf{97.88} & 0.7 \\
        \hspace{1em} Domain-level  & 95.03 & 93.49 & 98.33 & 95.85 & 0.4 & 99.79 & 99.49 & 96.03 & 97.72 & 0.4 \\
        \addlinespace
        Cookies                    & 91.76 & 96.96 & 92.75 & 94.81 & 0.4 & 93.89 & 44.31 & 100.00 & 61.41 & 0.4 \\
        Window Variables           & 95.47 & 96.82 & 96.25 & 96.54 & 0.4 & 99.78 & 95.65 & 98.214 & 96.92 & 0.4 \\
        \midrule
        Combined Classifier        & 95.28 & 96.93 & 94.34 & 95.62 & 0.5 & 99.22 & 80.42 & 97.46 & 88.12 & 0.5 \\
        Meta Classifier            & \textbf{96.75} & 96.82 & 97.22 & \textbf{97.02} & 0.5 & 99.72 & 96.26 & 98.10 & 97.17 & 0.5 \\
        \bottomrule[1.1pt]\\
    \end{tabular}

    \label{tab:combined-results}
\end{table*}

\subsection{Value Templates}
\label{subsec:methodology:value-templates}
\tool relies on the key observation that GA and sGA, when implemented as described in \Cref{sec:background}, produce the same semantic artifacts in the user's browser.
While publishers can rename query parameters, cookies, and window variables, the values these artifacts carry follow formats defined by GA, such as the client ID or the Google tag ID.
Therefore, instead of matching known names, we design \textit{Value Templates}: regex patterns that match the values of GA artifacts, regardless of the name they are stored under.
We construct Value Templates for each modality from client-side GA artifacts, i.e., artifacts collected on domains that send requests to known GA endpoints in our crawl (\Cref{subsec:methodology:browser-instrumentation}).
Value Templates used by \tool are listed in Table~\ref{tab:regexes}.

\para{Network requests}
We extract all query parameters from requests sent to known GA endpoints and rank them by frequency.
From the top 25 parameters, we remove those that carry time-based values, such as event timestamps, since they are not specific to GA.
For each remaining parameter, we iteratively design a regex pattern until a single, general pattern matches all values the parameter takes in our dataset.
For example, the parameter \texttt{cid} takes values such as \texttt{1834020512.1731024000} or \texttt{GA1.2.1834020512. 1731024000}, which we generalize to \texttt{(?:GA\bs d+\bs. \bs d+\bs.)?\bs d\{8,10\}\bs.17\bs d\{8,11\}}.

\para{Cookies}
We collect all first-party cookies on client-side GA domains, and use Cookiepedia~\cite{cookiepedia} to remove cookies that are not specific to GA, e.g., \texttt{\_gcl\_au}.
We then identify the unique types of GA cookies, such as \texttt{\_ga} and \texttt{\_ga\_<ID>}.
Since the values of these cookies follow different formats, we design a distinct regex pattern for each type.

\para{Window variables}
We collect all window variables on the same domains, excluding those that are functions, e.g., \texttt{onClick}.
We manually inspect GA-specific variables as specified by Google's documentation, such as \texttt{dataLayer}, \texttt{google\_tag\_data}, and \texttt{gaGlobal}, and design templates for the values they hold.

To avoid false positives, we apply each request template to the values of all other query parameters in our dataset, and refine any pattern that matches the values of a non-GA parameter, until no template matches values outside its own parameter.
This process results in 23 request, 5 cookie, and 10 window variable templates, shown in Table~\ref{tab:regexes}.

\noindent \textbf{Multiple Modalities.}
Using multiple modalities increases the robustness of \tool.
If one modality is obfuscated by the publisher, complementary modalities may still expose unaltered signals.
For instance, a publisher may encode network requests and hash cookie values, but may leave window variables set by Google’s script unchanged, enabling us to detect sGA through this alternate modality.

\subsection{Feature Extraction}
\label{subsec:methodology:feature-extraction}
We represent Value Template matches as binary feature vectors, which we use as input to train our classifiers.
For each modality, the feature vector has length $n$, equal to the number of Value Templates for that modality (\Cref{subsec:methodology:value-templates}), where each index is 1 if the corresponding template matches and 0 otherwise.

For network requests, we parse all query parameters in a given request URL and apply the request-based Value Template regexes to each parameter value.
We do not use the destination host or path as features, since they determine the training label (\Cref{subsec:methodology:classifier-training}).
For all first-party cookies on a domain, we serialize the cookie data into a string and apply the cookie-based Value Template regexes.
Finally, for JavaScript \texttt{window} variables, we call \texttt{JSON.stringify} on each variable and apply the window variable Value Template regexes.

\para{Feature Granularity}
A natural consequence of relying on features related to cookies and \texttt{window} variables is that they can only be computed at the domain level.
Thus, compared to one feature vector for cookies and \texttt{window} variables, we have one feature vector per network request, resulting in multiple binary feature vectors per domain.
While this granularity allows us to classify individual requests, it prevents us from combining features with other modalities (\eg for training a classifier that combines all features).
To this end, we also compute a single domain-level binary feature vector for network requests at the domain-level by taking a logical OR of all individual request-level feature vectors.

At the end of the feature extraction process, we end up with one feature vector for each modality at the domain level, and one feature vector for each network request.

\subsection{Classifier Training}
\label{subsec:methodology:classifier-training}
We train multiple classifiers to detect GA implementations using features from one or more modalities.
Since GA and sGA produce the same artifacts in the browser (\Cref{subsec:methodology:value-templates}), we train each classifier entirely on data from client-side GA, using a supervised bootstrapping strategy.

\para{Bootstrapping Training Labels}
We automatically annotate our dataset based on the destination of network requests observed in our crawl.
For any domain where we observe at least one network request sent to known GA endpoints, \textit{google-analytics.com} or \textit{analytics.google.com} (\Cref{subsec:background:google-analytics}), we label the domain as positive, and label all other domains as negative.
Similarly, for the request-level classifier, we label a request as positive if it is sent to a known GA endpoint, and negative otherwise.
Since negative domains include websites with other trackers and Google services, templates whose values also appear outside GA receive lower weights during training, so the classifiers learn which templates are specific to GA rather than relying on any single match.

\para{Classifier Architecture}
\tool uses Logistic Regression (LR), a supervised algorithm for binary classification.
We select LR over more complex models such as ensembles or neural networks for three reasons: its low inference cost suits real-time detection in a browser extension, its coefficients are interpretable, and those coefficients directly quantify each Value Template's contribution to detection.

\para{Training Multiple Classifiers}
Using domain-level features, we train six models.
First, we train a separate classifier for each individual modality: network requests, cookies, and window variables.
Second, we train a combined model using a single feature vector per domain that concatenates features from all three modalities.
Third, we train a meta-classifier that takes as input the probabilities from the three modality-specific models and learns a weight for each modality, enabling context-dependent weighting of modality signals based on their predictive importance.
Finally, we also train a request-level classifier using the individual network request feature vectors, so that we can identify which specific network requests, and therefore which endpoints and subdomains, carry GA data.
Together, these six models detect GA deployments purely by learning from the value-based patterns visible in client-side GA implementations.

\subsection{Deployment}
\label{subsec:methodology:deployment}
We deploy \tool as a Chrome Manifest V3 extension~\cite{sstguard}.
We export the trained models and replicate our feature extraction pipeline within the extension, which runs all classifiers (\Cref{subsec:methodology:classifier-training}) on the network requests, cookies, and window variables of each visited page.
Since cookies and window variables may be set asynchronously, the extension scans the page periodically, starting 500ms after the page load event.
For each page, the extension reports the output of each classifier, along with the specific requests, cookies, and window variables that matched the Value Templates.

\section{Validation \& Evaluation}
\label{sec:evaluation}
In this section, we describe the validation and evaluation process for \tool.
First, we explain our process for establishing ground truth for websites that use sGA (Section \ref{subsec:evaluation:ground-truth}).
Next, we explain our two-stage evaluation setup (Section \ref{subsec:evaluation:setup}) and discuss the performance of \tool (Section \ref{subsec:evaluation:results}).
Finally, we show that \tool's extension adds minimal overhead to page load time (Appendix \ref{subsec:evaluation:extension}).

\subsection{Ground Truth}\label{subsec:evaluation:ground-truth}
We establish ground truth with Google Tag Assistant (\Cref{subsubsec:background:tag-assistant}), which we only use for evaluation (\Cref{subsec:methodology:classifier-training}).
We crawl the Tranco top-10K websites (\Cref{subsec:methodology:browser-instrumentation}) with the extension loaded, refreshing each page three times to ensure the extension always loads, and record the destination URLs of all detected Google events.
We then remove destinations on known Google domains (\url{google-analytics.com}, \url{analytics.google.com}, \url{googleadservices.com}, \url{doubleclick.net}) and strip query parameters, leaving only server-side endpoints.
This process gives us \textbf{403 domains} that send Google events to non-Google destinations.

\subsection{Setup}
\label{subsec:evaluation:setup}
\noindent We evaluate \tool in two stages.

\para{Stage 1: Validation on training labels.}
We first validate each classifier on the bootstrapped labels (\Cref{subsec:methodology:classifier-training}) and select, per classifier, the probability threshold that maximizes F1 (Table~\ref{tab:combined-results}, Validation).

\para{Stage 2: Evaluation on sGA ground truth.}
We then evaluate on the 403 ground-truth domains (\Cref{subsec:evaluation:ground-truth}).
Since our classifiers detect \textit{any} GA implementation, we first remove client-side artifacts so that detections reflect sGA alone.
For network requests, we drop requests to GA endpoints (\textit{google-analytics.com}, \textit{analytics.google.com}), and recompute features on the remaining traffic.
Cookies and window variables cannot be attributed to the script that set them without taint tracking, so we instead evaluate these modalities only on domains with no client-side GA request.

\subsection{Detection Performance}
\label{subsec:evaluation:results}

\subsubsection{Stage 1}
On the bootstrapped labels, all domain-level classifiers achieve F1 above 94\%, with the \textit{Meta Classifier} performing best (97.02\% F1, 97.22\% recall; Table~\ref{tab:combined-results}, Validation).
The \textit{Request-level} classifier has lower precision (73.36\%), which we hypothesize is because requests to endpoints other than known GA endpoints carry GA-like information strongly enough to be flagged even at a high threshold.

\subsubsection{Stage 2}
On the sGA ground truth, \tool's classifiers generalize from client-side training data to server-side deployments (Table~\ref{tab:combined-results}, Evaluation).
The \textit{Request-level} classifier performs best, with 97.52\% recall and 97.88\% F1, followed closely by the \textit{Domain-level} (97.72\% F1), \textit{Meta} (97.17\%), and \textit{Window Variables} (96.92\%) classifiers.
The \textit{Request-level} classifier also improves sharply over Stage 1 (80.52\% to 97.88\% F1), confirming our hypothesis from earlier.
The \textit{Cookies} classifier is the exception: it detects every sGA domain (100\% recall) but with low precision (44.31\%), because GA cookies show up on pages that send no GA traffic at all (\Cref{subsec:evaluation:error-analysis}).
Across all modalities, \tool detects sGA on 401 of the 403 ground-truth domains, despite never being trained on sGA requests.

\subsection{Comparison with Prior Work}
\label{subsec:evaluation:prior-work}

\begin{table}[b]
\centering
\caption{Comparison with prior server-side detection approaches on the sGA ground truth (479 sites).}
\label{tab:prior-work-comparison}
\small
\begin{tabular}{@{}l l r@{}}
\toprule
\textbf{Approach} & \textbf{Features Used} & \textbf{Detected} \\
\midrule
Moti et al.~\cite{moti2025bitter}          & 36 Parameter names     & 116 (24.2\%) \\
Mertens et al.~\cite{mertens2026detecting} & GTM Containers & 248 (51.8\%) \\
\midrule
\textbf{\tool}                             & Value Templates          & \textbf{455 (95.0\%)} \\
\bottomrule
\end{tabular}
\end{table}

We compare \tool against prior approaches for detecting SST.
Fouad et al.~\cite{fouad2024devil} identify SST by comparing crawls from before and after the introduction of sGTM (March 2020 vs.\ May 2022).
Since their approach requires a historical crawl from before sGA was deployed, we cannot directly compare it to \tool.
We therefore compare \tool with Moti et al.~\cite{moti2025bitter}, who replace this crawl comparison with parameter matching, and with the concurrent work of Mertens et al.~\cite{mertens2026detecting}, who introduced a GTM-based detection mechanism.
To compare all three on the same websites, we run a fresh crawl of the Tranco top-10K domains.
We collect the data required by each approach: network requests for Moti et al.~\cite{moti2025bitter}, the output of the crawler provided by Mertens et al.~\cite{mertens2026detecting}, the outputs of the \tool Chrome extension, and ground truth from Google Tag Assistant~\cite{googleGoogleAssistant}.
\Cref{tab:prior-work-comparison} shows the results.
For all three approaches, we count a site as a successful detection only if the approach identifies the same endpoint as Tag Assistant.
\tool detects 455 of the 479 sites detected by Google Tag Assistant (95.0\%), compared to 116 (24.2\%) for Moti et al.\ and 248 (51.8\%) for Mertens et al.
Without the endpoint requirement, \tool detects 476 of the 479 sites (99.4\%) using any modality.
We discuss the reasons for these differences in more detail in \Cref{subsec:related:prior-server-side}.

\subsection{Error Analysis}
\label{subsec:evaluation:error-analysis}

We manually inspect the false positives and false negatives in Stage 2 to understand where \tool fails and why.
We find three main reasons.
First, we get false positives from other Google services (\eg Google Conversion Linker) that share identifiers with Google Analytics.
Second, we get false positives when GA state is present on a webpage but no GA traffic is sent.
Finally, we get false negatives from base64 encoded requests.
We provide the detailed analysis in Appendix~\ref{appendex:classification-errors}.

\para{Threshold selection}
Recall our observation from Stage 1, where the network request classifier showed lower precision (\Cref{subsec:evaluation:results}), suggesting that there are requests to non-GA endpoints with high similarity to GA requests.
We observe the opposite effect in Stage 2.
While validating false negatives in the Request-level classifier, we found that sGA requests on domains like \textit{gala.fr} were classified as negative at the 0.9 threshold.
These requests omit standard GA parameters like \texttt{cid}, \texttt{\_gid}, and \texttt{uapv}, and instead carry website-specific custom parameters (e.g., \texttt{ep.event\_name}) whose dynamic values our value templates cannot capture.
Together, these cases show that sGA spans a range of similarity to client-side GA: standard deployments score near 1, while customized deployments score lower even when they are functionally equivalent.
This is a systemic consequence of bootstrapping from client-side GA.
We therefore lower the Request-level threshold to 0.7, which extends coverage to customized deployments while maintaining 98.25\% precision on the ground truth.

\begin{figure*}[t]
    \centering
    \vspace{-0.5em}
    \begin{subfigure}[t]{0.32\linewidth}
        \centering
        \includegraphics[width=\linewidth]{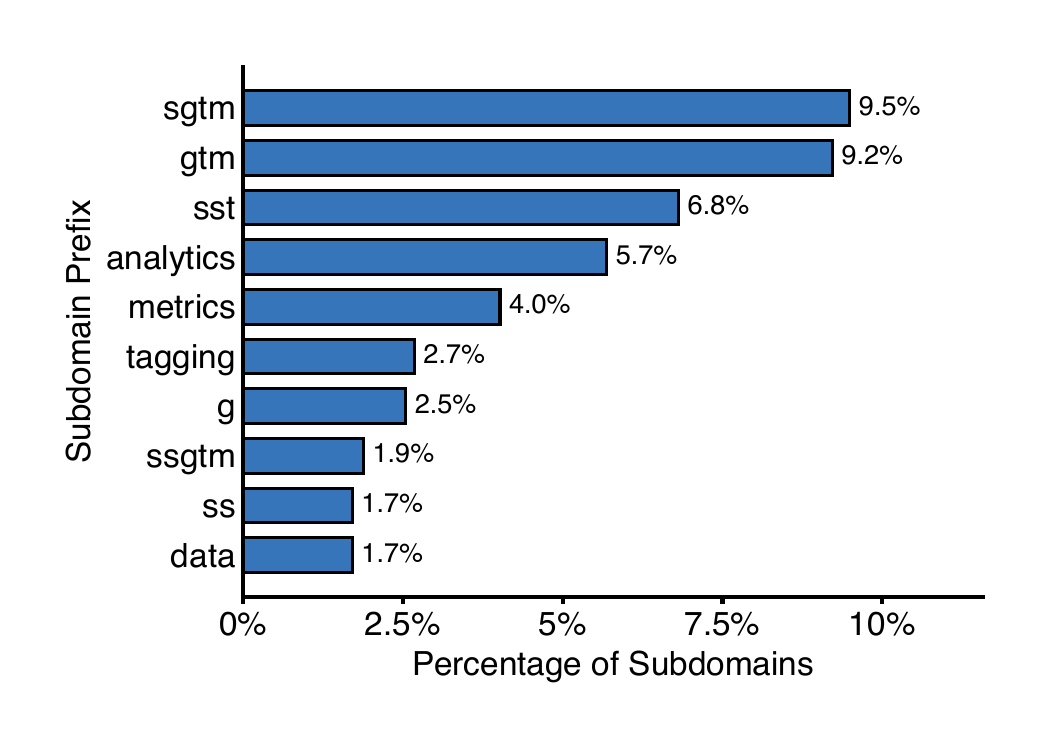}
        \caption{Most common subdomain prefixes (excludes path-based).}
        \label{fig:prefix-frequency}
    \end{subfigure}
    \hfill
    \begin{subfigure}[t]{0.32\linewidth}
        \centering
        \includegraphics[width=\linewidth]{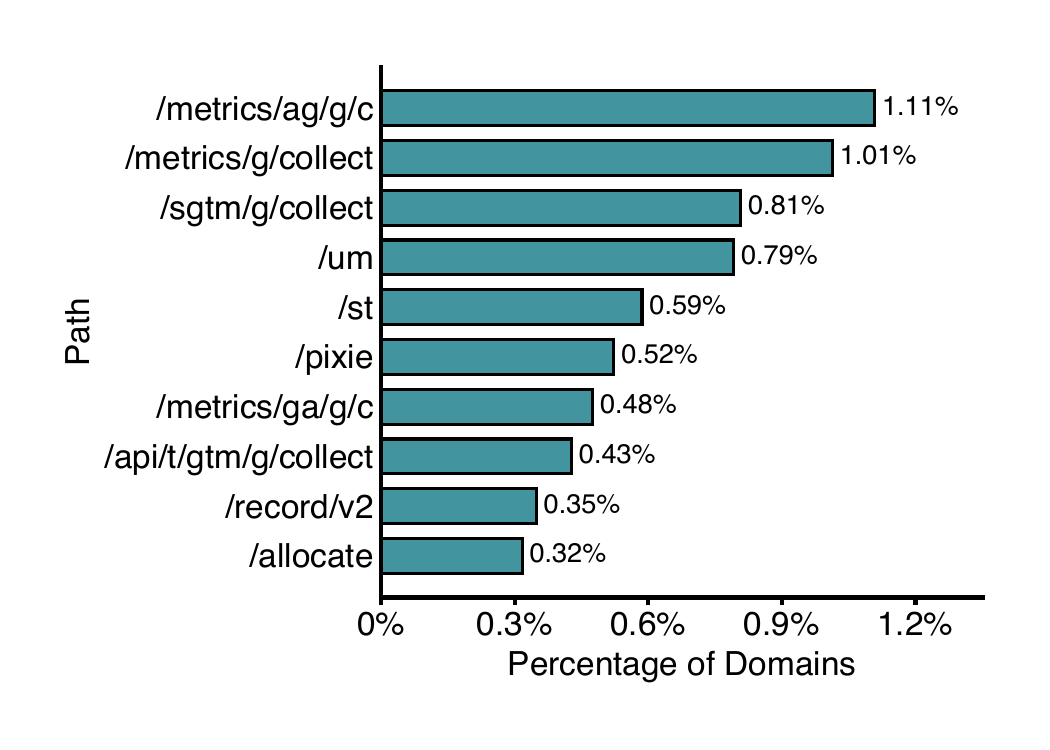}
        \caption{Top full request paths. Shows the most prevalent full request paths excluding \textit{\url{/g/collect}}.}
        \label{fig:top-paths}
    \end{subfigure}
    \hfill
    \begin{subfigure}[t]{0.32\linewidth}
        \centering
        \includegraphics[width=\linewidth]{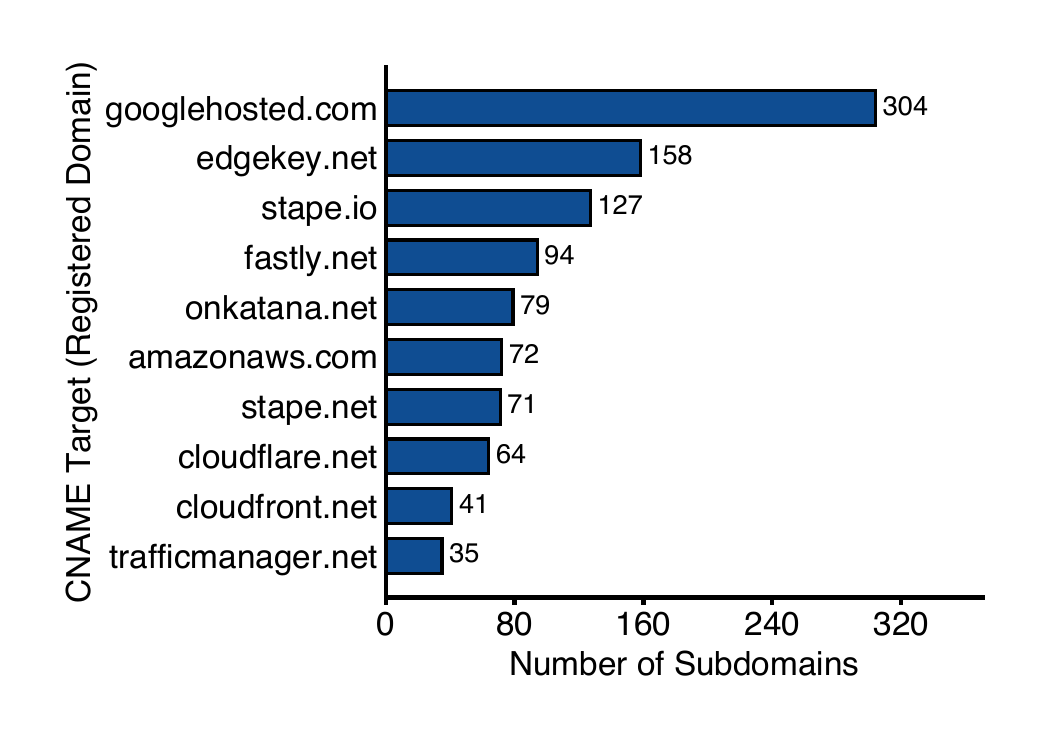}
        \caption{Third Party CNAME. Shows which domains are most frequently CNAME cloaked by publishers.}
        \label{fig:third-party-infra-providers}
    \end{subfigure}
    \caption{Characterizing sGA Implementations.}
    \label{fig:comprehensive-characterization}
\end{figure*}
\section{Characterizing SST In the Wild}
\label{sec:measurement}

We crawl the Tranco Top 150k domains to measure the prevalence of server-side Google Analytics in the wild.
For detecting sGA, we choose the \textit{Network Request - Request Level} classifier since it is the best performing model with the highest F1 score and precision (Table \ref{tab:combined-results}), allowing us to detect server-side tracking at the request level.
We first run the classifier on 128,222  domains, and then pass the output through EasyList~\cite{easylist} to remove all ad-tech domains which might use GA-related parameters, causing false positives.
Overall, we detect 40,199 sGA requests on 6,314 domains.

In this section, we dive deeper into the characteristics of these sGA implementations.
Our findings can help explain the current sGA ecosystem, answering questions like: where are typical sGA endpoints hosted? Which scripts trigger tracking requests? How customized are sGA requests? Can existing filterlists block sGA requests?

To this end, Sections~\ref{sec:measurement:implementation-patterns} through \ref{subsec:measurement:filterlist-analysis} discuss our insights from network requests.
Section~\ref{subsec:measurement:cookies-windows-analysis} discusses the variations in cookie and window variable naming.

\subsection{Implementation Patterns}
\label{sec:measurement:implementation-patterns}

We analyze the system architecture choices made by publishers when deploying sGA, specifically comparing subdomain-based routing to path-based routing of sGA requests.

\para{Routing strategies}
Our findings indicate that subdomain-based routing (\eg \textit{\url{sst.example.com}}) is the dominant deployment strategy, utilized by 81.59\% of the detected domains.
The remaining 18.4\% of domains opt for path-based sGA, routing tracking requests through an endpoint on the primary domain (\eg \textit{\url{www.example.com/track}}).
Since these endpoints share the DNS records of the main website, our network level analysis does not apply to them.

\para{Subdomain naming conventions}
Among the websites using distinct subdomains, we observed a variety of different prefixes.
Figure \ref{fig:prefix-frequency} shows our findings.
The top three most prevalent prefixes are \texttt{sgtm} (9.49\%), \texttt{gtm} (9.22\%), and \texttt{sst} (6.81\%).
Other frequently observed subdomains include \texttt{analytics}, \texttt{metrics}, and \texttt{tagging}.

\para{Path analysis}
We further examined the request paths used to send tracking data to server-side.
The default Google Analytics endpoint path, \textit{/g/collect} remains the most common by a significant margin, appearing on 4,542 domains (71.94\% of all sGA endpoints).
In fact, on 5,049 domains (79.97\% of all sGA endpoints), we observe the last path segment to be \textit{/collect}.
We also observe non-standard endpoints like \textit{/metrics/ag/g/c} (1.11\%) and \textit{/metrics/g/collect} (1.01\%).
Figure~\ref{fig:top-paths} shows the top 10 most frequently occurring paths, excluding the default Google path \textit{/g/collect}.
These results demonstrate low adoption rates of custom paths, indicating that fewer developers are utilizing the full customization offered by sGA (Section \ref{subsec:background:google-analytics}) and instead defaulting to subdomain-based routing.

\begin{takeaway}
Most publishers (81.59\%) route sGA through subdomains with descriptive prefixes such as \texttt{sgtm}, \texttt{gtm}, and \texttt{sst}, and 71.94\% keep the default \texttt{/g/collect} path.
These prefixes and paths can be added to filterlists, but the remaining 20.03\% of publishers use a long tail of custom paths that per-site rules cannot keep up with.
\end{takeaway}

\subsection{Network Level Analysis}
\label{sec:measurement:network-level-analysis}

To understand how publishers integrate sGA into their infrastructure, we analyze the network architecture of subdomain-based deployments (e.g., \texttt{\url{sst.example.com}}).
For each detected sGA endpoint, we extract the Fully Qualified Domain Name (FQDN) and perform three analytical steps:
First, we classify the DNS resolution output, by querying both the publisher's main domain and the sGA subdomain using \texttt{dnspython} with the resolver set to Cloudflare (1.1.1.1).
We categorize endpoints based on whether they resolve via CNAME records (pointing to third-party infrastructure) or direct A/AAAA records (indicating first-party IP ownership).
Then, we map IP addresses to Autonomous System Numbers (ASNs) using Team Cymru's IP-to-ASN service to identify the organizational ownership of the hosting infrastructure~\cite{teamcymruMappingService}.
Finally, we cross-reference these ASNs against CAIDA's AS-to-Organization dataset~\cite{caidaOrganizationsMappings} to determine whether the sGA infrastructure differs from that of the publisher's main website.
This reveals whether the publishers host tracking on the same infrastructure as their primary content, or delegate to external providers.
We explain our findings in this section, summarizing them in Figure~\ref{fig:sankey}.

\para{CNAME Cloaking Prevalence}
We find that 1,365 domains (26.49\% of the 5,152 domains) employ CNAME cloaking.
Of these, 2.64\% (36 domains) resolve to first-party domains, most likely used for internal routing and management.
The rest directly resolve to a third-party.
The remaining 73.51\% (3,787 domains) use A/AAAA records.

\para{Third-Party Infrastructure Providers}
Analysis of the 1,329 third-party CNAME subdomains reveals a highly consolidated infrastructure landscape, with the top ten providers accounting for 74.94\% of all CNAME cloaked subdomains (Figure~\ref{fig:third-party-infra-providers}).
Google dominates sGA hosting via \textit{googlehosted.com} (319 endpoints), indicating that these publishers chose the default GCP Cloud Run option.
Stape.io ranks second with 199 endpoints combined across \textit{stape.io} (133) and \textit{stape.net} (66).
Amazon Web Services (AWS) follows as the third-largest provider with 120 endpoints split between general cloud infrastructure (\textit{amazonaws.com}, 79) and their CDN (\textit{cloudfront.net}, 41).
CDN and edge providers constitute another frequent occurrence: Akamai (\textit{edgekey.net}) hosts 124 endpoints, Fastly (\textit{fastly.net}) hosts 90 endpoints, and Cloudflare (\textit{cloudflare.net}) accounts for 77 endpoints.

\para{Infrastructure Ownership Analysis}
We now examine deployments that do not rely on CNAME cloaking, \ie, domains that resolve to A/AAAA records.
For these 3,787 (73.51\%) subdomains, we determine infrastructure ownership as explained earlier.
We categorize these as either \emph{same infrastructure} (sGA subdomain shares ASN/organization with the main domain) or \emph{infrastructure shift} (delegated to external hosting providers).
Out of 3,787 analyzed subdomains, 3,132 (82.70\%) represent an \emph{infrastructure shift}, where publishers delegate sGA hosting to external providers rather than using their primary infrastructure.
Among these shifted deployments, Google dominates with 2,581 endpoints (82.41\% of all shifted subdomains), split between Google Cloud Platform (1,646) and Google (935).
Amazon ranks the second with 215 endpoints (Amazon-02 and Amazon-AES), followed by AS12876 (112), Cloudflare (42), Microsoft (25), and others.

\begin{takeaway}
73.51\% of sGA subdomains use direct A/AAAA records instead of CNAME cloaking, which makes them undetectable by CNAME-based defenses such as Safari's.
82.70\% of these are hosted on infrastructure different from the publisher's main website, mostly on Google Cloud Run, the default sGTM option.
\end{takeaway}

\subsection{Script Analysis}
\label{subsec:measurement:script-analysis}

\para{Initiator Domain Analysis}
Our analysis of network request initiators reveals that \textit{www.googletagmanager.com} remains the most prevalent source, initiating tracking requests for 31.5\% of sites in our dataset.
Despite Google's recommendation to serve scripts through the first-party endpoints for improved resilience, loading from Google's domain leaves these implementations vulnerable to blocking via standard filterlists like EasyPrivacy~\cite{easyprivacy}.
Other significant third-party initiators include \textit{www.datadoghq-browser-agent.com} (1.2\%) and \textit{img1.wsimg.com } (1.2\%).
We also observe a significant portion of ``Unknown'' initiators, typically representing requests originating from scripts that the Chrome DevTools Protocol (CDP) could not attribute to a specific script stack~\cite{CDPnetworkDomain}.
The high prevalence of standard third-party initiators suggests that while the \textit{data destination} has shifted to the server-side, the \textit{initialization logic} for over a third of the ecosystem remains anchored to known third-party domains, providing an easy leverage point for existing client-side defenses.

\para{Initiator Script Analysis}
Regarding the specific scripts initiating these requests, we find that \texttt{gtag.js} is the most prevalent, acting as the final executor for 34.7\% of total requests.
The second most common initiator script is the primary HTML document itself (\texttt{root\_index}), which accounts for 13.2\% of requests.
This indicates that tracking code is called directly from the HTML, instead of other Javascript files.

\begin{takeaway}
\texttt{gtag.js} initiates more sGA requests (34.7\%) than \texttt{googletagmanager.com} (31.5\%), indicating that only a few publishers serve it from first-party infrastructure.
\end{takeaway}

\subsection{Payload Analysis}
\label{subsec:measurement:payload-analysis}

Next, we analyze the query parameters present in URLs identified as SST requests.
We categorize observed keys into three distinct groups: \textit{standard} keys (e.g., \texttt{tid}, \texttt{cid}), \textit{SST-specific} keys added by the server-side container (e.g., \texttt{sst.tft}), and \textit{non-standard} keys consisting of custom event parameters and user properties.
Standard keys appear in 97.3\% of all requests, while non-standard custom parameters are present in 96.6\%.
Notably, SST-specific parameters are only observed in 76.5\% of requests.

\para{Standard Keys}
The prevalence of standard parameters is high; each of the top 10 keys appear in over 93\% of all requests.
The most frequent key, \texttt{gtm} appears in 95.7\% requests.
Other frequent parameters include \texttt{tag\_exp} (95.5\%), and \texttt{gcd} (95.1\%), which carries Google's Consent Signals.

\para{SST Keys}
Since \texttt{sst.} parameters are specific to server-side containers, we expect them in nearly every sGA request.
Unexpectedly, only 76.5\% of requests contain parameters these parameters, which suggests that the remaining requests are sent to an intermediary that does not use the default server-side container configuration.
Among these requests, we find that most implementations use path-based routing (59.06\%).
We also observe that only 13.60\% requests use the \textit{/g/collect} endpoint, compared to 71.94\% of all sGA domains.
Requests using SST keys also contain a high number of standard keys---the top 10 most frequent standard keys ranging from 71\% to 82\% of the requests.
In contrast, the top 10 non-standard keys range from 7\% to 50\% of the requests
This indicates that these 1,615 (25.58\% of all domains) domains are heavily customizing their sGA implementations and overriding default behaviors.

\para{Non-Standard Keys}
Non-standard keys represent site-specific data transmission.
The most frequent custom keys are \texttt{gap.gtb} (34.7\%) and \texttt{dma\_cps} (34.4\%), which is the conventional parameter name for conveying the \textit{Digital Markets Act Consent Parameter Settings}.
We also observe high prevalence for \texttt{ir} (29.0\%), and \texttt{ep.event\_id} (26.0\%), a unique identifier likely used to de-duplicate events (\eg for GA4 events or Meta's Conversions API) \cite{gaDuplicateEventsBlog2026, facebookIntegrateConversionsWithSGTM}.
Finally, \texttt{ec\_mode} (24.2\%) indicates the use of \textit{Enhanced Conversions}, which often involves the server-side hashing and transmission of user-provided data such as email addresses or phone numbers.

\begin{takeaway}
25.58\% of sGA domains heavily customize their requests and override default sGA behaviors.
Non-standard keys such as \texttt{ep.event\_id}, \texttt{ep.fb\_event\_name}, and \texttt{ec\_mode} suggest that publishers share event data with other platforms and transmit hashed user data through sGA.
\end{takeaway}

\subsection{Filterlists}
\label{subsec:measurement:filterlist-analysis}
To see how much of the sGA we detect is already blocked by filterlists, we apply EasyPrivacy~\cite{easyprivacy} to the 40,199 sGA requests we identify in the wild.
We match each request using the \texttt{adblock-rs} engine~\cite{docsAdblockRust} (v0.12.1), and record the rule that blocks it.

EasyPrivacy blocks 37,585 requests (93.50\%).
This high coverage is achieved through specific path-based rules targeting (1) standard GA4 endpoint: generic \texttt{/collect} path patterns (with 26,533 matches), (2) the GA4 Measurement Protocol signature \texttt{?v=2\&tid=G-} with first-party targeting (with 4,907 matches), and (3) Server-Side Tag Manager specific parameters such as \texttt{\&sst.sw\_exp=} (with 2,231 matches).
Notably, the rule \texttt{?v=2\&tid=G-\$\textasciitilde third-party} explicitly targets first-party traffic (the ``\textasciitilde'' modifier negates the third-party option), indicating that EasyPrivacy has specific rules to detect sGA deployments that use first-party domains with default GA4 protocol signatures.
Further, the presence of rules for specific sGTM state parameters (\texttt{\&sst.sw\_exp=}) confirms that EasyPrivacy maintainers have explicitly added signatures for server-side container deployments.
Finally, we also observe rules for blocking specific server-side endpoints like \textit{mstm.motorsport.com}, increasing coverage.

However, this 93.50\% coverage represents a best-case scenario for filterlists and masks a dependence on only a few rules.
The blocked traffic relies on two assumptions: (1) that publishers use the default \texttt{/g/collect} endpoint path, and (2) that they transmit unmodified GA4 parameters in order (\texttt{v=2}, \texttt{tid=G-}).
As established in \Cref{sec:measurement:implementation-patterns}, while 71.94\% of sGA domains use the default \texttt{/g/collect} path, the remaining domains use custom paths that these rules do not cover.
For example, requests to \textit{\url{https://www.nj.betmgm.com/reverse-proxy}} and \textit{\url{https://ss-tracking.pimkie.fr/czi4bu}} were not blocked because of such customizations.
Overall, 166 domains (2.62\%) have no sGA requests blocked.
Further, since publishers can modify, encode, or completely replace query parameters (\Cref{subsubsec:sga-customization}), the \texttt{tid=G-} and \texttt{sst.*} signatures are ineffective against obfuscated deployments.
For example, the \texttt{base64} encoded requests on \textit{themeisle.com}, \textit{shapeways.com}, \textit{comparitech.com}, and \textit{measureschool.com} are not blocked by EasyPrivacy, while \tool detects sGA on all four websites.
We reported these gaps to the EasyPrivacy maintainers, who have since added rules based on our findings (\Cref{subsec:discussion:impact}).

\begin{takeaway}
EasyPrivacy blocks 93.50\% of sGA requests, but this coverage depends on a few rules that assume the default \texttt{/g/collect} path and unmodified GA4 parameters.
Customized and encoded deployments evade these rules, leaving 166 domains with no requests blocked.
\end{takeaway}

\subsection{Cookies and Window Variables}
\label{subsec:measurement:cookies-windows-analysis}
To understand customized naming of cookies and window variables used in the wild, we apply our value templates to the detected sGA domains.

\para{Cookies}
In our dataset, we find cookies on 6,306 domains with SST.
We observe a few variations of cookie names apart from the default \texttt{\_ga} cookie name.
First, our value template for \texttt{\_ga} matches the \texttt{\_gid} cookie on 941 domains (14.9\%), which is also a first party cookie set by some Google Analytics implementations to track visits across multiple days.
Other than \texttt{\_gid}, we observe some websites (e.g., \textit{boconcept.com}, \textit{boisedev.com}, \textit{dnv.com}, \textit{bowerswilkins.com}) to use custom cookie names for \texttt{\_ga} cookie, resulting in matches like \texttt{csparkW\_ga} (37 domains), \texttt{\_ga\_backup} (20 domains), \texttt{ga4\_ga} (5 domains), and \texttt{local\_ga} (5 domains).
For the \texttt{\_ga\_X} type cookie, we find some instances of cookie renaming.
For instance, we find cookie names like \texttt{\_ga\_12345} (54 domains), \texttt{\_ga\_1234} (37 domains), and \texttt{\_ga\_123456789} (31 domains).
\para{Window Variables}
We observe a lot of diversity in window variable naming.
While the default \texttt{dataLayer} is the most commonly used variable name, found on 4,744 domains (75.13\% of domains on which we successfully collected window variables), we find 260 other variations.
For example, we observe the variable names \textit{\_tccInternal} on 120 domains (1.9\%), and \textit{\_sGtmDataLayer}, \textit{\_analyticsDataLayer} and \textit{\_wGtmDataLayer}, each on 119 domains (1.88\%).
Interestingly, we also observe other tag managers and tag support tools used by developers.
For instance, we see the \texttt{\_dlo\_observer} variable on 40 domains (0.63\%).
This is from the FullStory DataLayer Observer~\cite{Fullstorydevfullstorydatalayerobserver}, a JS library for analytics and tag integration.
We also notice the \texttt{MatomoTagManager} variable used by the server-side tag manager Matomo Tag Manager~\cite{matomoListFeatures} on 18 domains (0.29\%).
This demonstrates that our value templates were able to capture a wide array of implementations, including those which do not rely on Google Tag Manager.

Similarly, the variables \texttt{google\_tag\_data} and \texttt{google\_tag\_manager} are detected by our value templates on 5,099 and 4,414 domains (80.76\% and 69.90\%) respectively.
We also matched their other variations such as \texttt{googletag}, \texttt{utag}, \texttt{ga4\_measurement\_id} and \texttt{utag\_data}.

\begin{takeaway}
Most publishers keep default cookie names, but rename window variables heavily, with 260 variations of \texttt{dataLayer} alone being detected by \tool.
Since \tool matches values rather than names, it detects these renamed variables, including those set by tag managers other than GTM.
\end{takeaway}

\section{Related Work}
\label{subsec:related-works}

\subsection{Client-side tracking research}
Prior work has largely studied tracking that executes in the browser~\cite{vekaria2025sok}.
As platforms increasingly embed pixels and tags in the first-party context~\cite{munir2023cookiegraph, jazlan2026tracking}, studies have examined what Meta's pixel collects, including PII~\cite{bekos2023hitchhiker, bekos2025piixel}, how its configuration varies across websites~\cite{ghani2026pixelconfig, kieserman2025tracker}, and the privacy implications of Google Tag Manager configurations~\cite{mertens2025you, moti2025bitter} and templates~\cite{mertens2026detecting}.
Privacy tools counter such tracking with community-maintained filterlists~\cite{easylist, easyprivacy}, which are inherently reactive and miss over a quarter of trackers~\cite{fouad2020missedfilterlistsdetecting}.
Trackers can further evade these defenses through CNAME cloaking, aliasing a first-party subdomain to a third-party tracker, which prior work detects by resolving the DNS indirection~\cite{dao2021cname, dimova2021cname}.
These approaches largely miss server-side tracking such as sGA.
In our measurement, roughly 21\% of sGA domains are CNAME cloaked, and these resolve not to a known tracker but to generic cloud infrastructure.

\subsection{Prior Server-side Tracking Research}
\label{subsec:related:prior-server-side}
El Fraihi et al.~\cite{elfraihi2024serverside} compared Meta's client-side pixel against its server-side CAPI, finding comparable effectiveness but 35\% lower accuracy for CAPI.
Beyond measurement, existing detection approaches rely on artifacts that the publisher controls and can change: request parameter names~\cite{fouad2024devil, moti2025bitter} or a specific GTM deployment~\cite{mertens2026detecting}.
We compare \tool against both and show that customization, obfuscation, or a different deployment pattern is enough to evade them.

\subsubsection{Network-level heuristics}
Fouad et al.~\cite{fouad2024devil} identify server-side tracking deployed with GTM by comparing crawls from before and after the introduction of sGTM (March 2020 vs. May 2022), and confirming that the new subdomains belong to a different organization than the website.
Moti et al.~\cite{moti2025bitter} use parameter matching: they collect the URL parameters shared across GTM-initiated requests (36 in total), search for them in first-party traffic, and validate hits by checking whether the subdomain's IP belongs to a different organization.
This design rests on three assumptions: (i) requests keep their default parameter names, (ii) GTM initiates the request, and (iii) the endpoint is a distinct first-party subdomain.
sGA breaks all three (\Cref{subsec:background:sga-deployment}): publishers can strip or encode parameters, call gtag.js directly, or host endpoints on a path or behind a load balancer.
The authors acknowledge that their method may not generalize to more customized uses of SST~\cite{moti2025bitter}.
We replicate their approach on our ground truth with the 36 parameters obtained from the authors, and identify 116 of 479 domains (\textbf{24.2\% recall}), compared to \textbf{95\%} for \tool.

\subsubsection{GTM Container-based detection}
Concurrently to our work, Mertens et al.~\cite{mertens2026detecting} instead parse GTM Containers for Collector Tags configured with a server container URL, detecting 3,095 candidate domains from 80.6K websites.
This restricts detection to one deployment method: a GTM Container with one of three Collector Tags.
It therefore misses sGA deployed with \texttt{gtag.js} alone, custom collectors, and other tag managers, all of which \tool detects through the values of client-side artifacts.
On the same set of websites, their artifacts reach \textbf{51.8\%} recall, compared to \textbf{95\%} on the network request modality for \tool and \textbf{99.4\%} using all modalities.

We identify four cases in which Mertens et al.~\cite{mertens2026detecting} fail to detect sGA, on websites that:
(i) run sGA with \texttt{gtag.js} alone, without GTM;
(ii) host the sGA endpoint on a path of the canonical domain;
(iii) configure the endpoint through the \texttt{transport\_url} setting (\Cref{subsec:background:sga-deployment});
and (iv) set the endpoint URL from a JavaScript variable.

\section{Discussion}
\label{sec:discussion}

\subsection{Robustness to Obfuscation}
\label{subsec:discussion:robustness}
\tool detects multiple websites that obfuscate their sGA requests to evade filterlists.
These websites use Stape.io's Advanced GA tag~\cite{stapeAvoidBlocking, stapeLoadContainer}, which strips all query parameters from outgoing sGA requests and \texttt{base64} encodes them.
For example, on \textit{shapeways.com}, an sGA request to \textit{gtm.shapeways.com} (Table~\ref{tab:ssga-urls}) decodes to \textit{/g/collect?v=2\&tid=G-MXZEPNTKS8...}, following the GA4 endpoint convention.
To measure the prevalence of this pattern, we search for the encoded segment \texttt{L2cvY29sbGVjdD92PTImdGlkPUct} in our dataset, which decodes to \texttt{/g/collect?v=2\&tid=G-}.
We find 223 domains using this pattern, demonstrating active attempts to evade privacy tools like EasyPrivacy~\cite{easyprivacy}.

Since these requests carry no query parameters, the request-level classifier detects only 7 of these domains (those that also send un-encoded requests alongside encoded ones).
In contrast, our cookie and window variable Value Templates detect sGA on 194 and 172 domains respectively, totaling 216 out of 223 domains (\textbf{96.86\%}).
This shows that obfuscating one modality is not enough to evade \tool.

Window variables are the hardest of these modalities to obfuscate.
Prior work has demonstrated obfuscation of network requests~\cite{sjosten2020filter, siby2022webgraph, munir2024purl}, cookies~\cite{munir2023cookiegraph}, and the JavaScript code delivered to a website~\cite{sarker2020hiding, chen2021detecting} in web tracking, making detection based on these signals alone unreliable.
However, regardless of how the code is delivered, it must interact with the browser once executed, leaving artifacts in the page's JavaScript environment.
For sGA, these artifacts are the window variables created by \texttt{gtag.js} and \texttt{gtm.js}, such as \texttt{dataLayer}, \texttt{google\_tag\_data}, and \texttt{gtag}.
These variables are the interfaces through which publishers configure and send events to GA, so the values they hold must follow a structure that GA expects, even if a publisher renames them.
Even custom implementations that collect events in the browser must store the information GA requires to attribute these events, such as the Google tag ID (\Cref{subsec:background:sga-deployment}), in the page's JavaScript environment.
Since millions of websites integrate GA and GTM through these interfaces~\cite{WebAlmanac.2025.ThirdParties, mertens2026detecting}, changing this structure would require publishers to re-implement their tracking, creating strong inertia against such changes.
Therefore, while window variables can also be obfuscated with highly customized implementations, they are likely to remain a reliable signal for \tool when network and cookie signals are not.

\subsection{Privacy Concerns with SST}
\label{subsec:discussion:privacy-concerns}
First, server-side tracking takes away transparency and control from users.
Since SST disguises tracking traffic as first-party, users can no longer identify which trackers are embedded on a page, or what information is shared with them.
For instance, with client-side tracking, users can identify requests going to third-party tracking endpoints, like \textit{analytics.google.com} or \textit{facebook.com/tr}, and block them using ad blockers.
In contrast, a single server-side request can carry the same information to a first-party SST endpoint, from where it can be further shared with any number of tracking platforms.

Second, server-side tracking shifts cross-site tracking to identifiers that users cannot control.
Since server-side trackers cannot rely on third-party cookies, they use first-party cookies combined with other identifiers, such as hashed email addresses for advanced matching tools (\eg Enhanced Conversions, \Cref{subsec:measurement:payload-analysis}) or IP address and user agent (\ie fingerprinting).
Users can block third-party cookies, but they cannot block first-party cookies without breaking most websites, and have no control over their email address being shared or over fingerprinting.

Finally, while SST gives publishers control over what data is shared with trackers, this control is not unique to SST.
Client-side trackers already offer publishers extensive configuration options~\cite{kieserman2025tracker, ghani2026pixelconfig}, and SST offers similar, if somewhat broader, configuration on the server.
Thus, SST provides little additional privacy benefit, while taking away transparency and control from users.

\subsection{Impact of \tool}
\label{subsec:discussion:impact}
We reported our findings from \Cref{sec:measurement} to the maintainers of EasyPrivacy~\cite{easyprivacy}.
They integrated the most common subdomain names and paths (\Cref{sec:measurement:implementation-patterns}) to provide the highest coverage with the fewest rules.
These changes can be seen in the cited commits~\cite{easyprivacy2026-commit-40e35ea, easyprivacy2026-commit-fa4c7d3, easyprivacy2026-commit-2331cb8, easyprivacy2026-commit-365b772, easyprivacy2026-commit-da0c69a, easyprivacy2026-commit-0da4049, easyprivacy2026-commit-663da24, easyprivacy2026-commit-96f8851, easyprivacy2026-commit-42e7847, easyprivacy2026-commit-ec812ae, easyprivacy2026-commit-e2b53d3, easyprivacy2026-commit-d5d631f, easyprivacy2026-commit-1a11ea3}.
In addition, we open source \tool as a Chrome extension with this paper~\cite{sstguard} so that new and emerging patterns can be detected before they are added to a filterlist.

\subsection{Limitations and Future Work}
\label{subsec:discussion:limitations}
Our approach has three main limitations.
First, our detection relies on the semantic similarity between client-side and server-side implementations.
If Google were to restructure sGA such that transmitted values no longer resemble client-side patterns, our value templates would no longer match.
However, such a change would also break existing GA4 integrations across millions of websites.

Second, our ground truth relies on Google Tag Assistant, which only captures implementations that use Google's \texttt{gtag} library.
Custom sGA implementations that use the Measurement Protocol (MP) directly, such as those offered by Freshpaint~\cite{freshpaintGoogleAnalytics}, are not captured.
Our results therefore present a lower bound on the prevalence of sGA.
However, \tool can still detect MP implementations that collect GA values in the browser.
For example, \tool detects Mountain, an analytics pixel that collects the website's GA configuration and reports to Google Analytics via the MP~\cite{mountainGA}, although we exclude such known third parties from our results.

Third, \tool does not currently generalize to other server-side trackers.
While the semantic invariance we rely on is not specific to GA (\eg Meta's \texttt{\_fbp} cookie and standard event names are required parameters for its Conversions API), extending \tool to other trackers requires more than new value templates.
Meta, TikTok, Snapchat, and Reddit do not offer a debugging tool equivalent to Tag Assistant, leaving no reliable source for ground truth.
Further, our approach assumes that trackers leave identifiable values on the client side.
\tool is robust to customizations that change how these values are transmitted, such as renamed keys, rerouted endpoints, or encoded payloads, but not to deployments where the values themselves never reach the browser.
For example, even with a reliable source of ground truth for Meta and corresponding value templates, a Meta CAPI deployment that maps custom event names to Meta's standard schema on the server-side would still evade detection, since the values observed on the client side would not match the templates.
Detecting such deployments would likely require cross-layer approaches, such as taint tracking from DOM events to network traffic, which we leave to future work.

\section{Conclusion}
\label{sec:conclusion}

In this paper, we propose \tool, a novel approach that detects sGA using value templates, which match the values of Google Analytics artifacts rather than their names.
\tool applies value templates across network requests, cookies, and JavaScript \texttt{window} variables, which allows it to detect sGA even when endpoints are customized or query parameters are obfuscated.
On sGA ground truth collected using Google Tag Assistant, \tool detects the sGA endpoint on 95.0\% of websites, compared to 51.8\% for concurrent work and 24.2\% for prior work.
Applied to the Tranco top-150K, \tool detects sGA on \totalDetected websites.
We find that publishers customize their sGA deployments using path-based endpoints, non-standard paths, renamed variables, and base64-encoded query parameters.
We reported our findings to the maintainers of EasyPrivacy, resulting in 26 rules that improve its coverage of sGA.
Overall, \tool and its value templates provide a foundation for future research on server-side tracking, including on other tracking platforms.

\section*{Acknowledgments}
We would like to thank Akhil Guntur for his contributions to the project.

\bibliographystyle{IEEEtranS}

\bibliography{ref}

@misc{mountainGA,
  author       = {MNTN},
  title        = {Connect {Google Analytics}},
  howpublished = {\url{https://help.mountain.com/en/articles/11534853-connect-google-analytics}},
  note         = {Accessed: 2026-11-14}
}

@misc{openaiCAPI,
	author = {},
	title = {{C}onversions {A}{P}{I} – {A}ds | {O}pen{A}{I} {D}evelopers --- developers.openai.com},
	howpublished = {\url{https://developers.openai.com/ads/conversions-api}},
	year = {2026},
	note = {[Accessed 17-11-2026]},
}

@misc{sstguard,
	author = {Muhammad Jazlan and Alexander Gamero-Garrido and Zubair Shafiq and Yash Vekaria},
	title = {{SST-Guard}},
	howpublished = {\url{https://github.com/jazlan01/sst-guard}},
	year = {2026},
}

@inproceedings{sarker2020hiding,
  title={Hiding in plain site: Detecting javascript obfuscation through concealed browser api usage},
  author={Sarker, Shaown and Jueckstock, Jordan and Kapravelos, Alexandros},
  booktitle={Proceedings of the ACM internet measurement conference},
  pages={648--661},
  year={2020}
}

@inproceedings{chen2021detecting,
  title={Detecting filter list evasion with event-loop-turn granularity javascript signatures},
  author={Chen, Quan and Snyder, Peter and Livshits, Ben and Kapravelos, Alexandros},
  booktitle={2021 IEEE Symposium on Security and Privacy (SP)},
  pages={1715--1729},
  year={2021},
  organization={IEEE}
}

@misc{easyprivacy2026-commit-40e35ea,
  author       = {{ryanbr}},
  title        = {{EasyPrivacy Commit 40e35ea}},
  year         = {2026},
  month        = nov,
  day          = {4},
  howpublished = {\url{https://github.com/easylist/easylist/commit/40e35eac1e2afff8ff6b22ea713aeec19d6d40f9}},
  note         = {Adds the filter \texttt{/g/collect?} to \texttt{easyprivacy\_general.txt}. Accessed: 2026-11-09}
}

@misc{easyprivacy2026-commit-fa4c7d3,
  author       = {{ryanbr}},
  title        = {{EasyPrivacy Commit fa4c7d3}},
  year         = {2026},
  month        = nov,
  day          = {8},
  howpublished = {\url{https://github.com/easylist/easylist/commit/fa4c7d3de3fcd997adc3e0df6f115401f4828d5d}},
  note         = {Adds the filters \texttt{/api/t/gtm/*}, \texttt{/metrics/ga/*}, and \texttt{/sgtm/g/*} to \texttt{easyprivacy\_general.txt}. Accessed: 2026-11-09}
}

@misc{easyprivacy2026-commit-2331cb8,
  author       = {{ryanbr}},
  title        = {{EasyPrivacy Commit 2331cb8}},
  year         = {2026},
  month        = nov,
  day          = {8},
  howpublished = {\url{https://github.com/easylist/easylist/commit/2331cb816129ab40f7ed57c480902fd0f82b7eb1}},
  note         = {Adds the filters \texttt{/tms?id=} and \texttt{://sgtm.} to \texttt{easyprivacy\_general.txt}. Accessed: 2026-11-09}
}

@misc{easyprivacy2026-commit-365b772,
  author       = {{ryanbr}},
  title        = {{EasyPrivacy Commit 365b772}},
  year         = {2026},
  month        = nov,
  day          = {8},
  howpublished = {\url{https://github.com/easylist/easylist/commit/365b7721143c7e8e676ae1463fd7a9908e36a440}},
  note         = {Adds the filters \texttt{/collect/metric|} and \texttt{/v1/events/publish|} to \texttt{easyprivacy\_general.txt}. Accessed: 2026-11-09}
}

@misc{easyprivacy2026-commit-da0c69a,
  author       = {{ryanbr}},
  title        = {{EasyPrivacy Commit da0c69a}},
  year         = {2026},
  month        = nov,
  day          = {8},
  howpublished = {\url{https://github.com/easylist/easylist/commit/da0c69a45a004a3d775bbcc80e92a09723dc4876}},
  note         = {Adds the filters \texttt{/cohesion/xs2.html} to \texttt{easyprivacy\_general.txt}; \texttt{||psychcentral.com/api/drivers/} to \texttt{easyprivacy\_specific.txt}. Accessed: 2026-11-09}
}

@misc{easyprivacy2026-commit-0da4049,
  author       = {{ryanbr}},
  title        = {{EasyPrivacy Commit 0da4049}},
  year         = {2026},
  month        = nov,
  day          = {8},
  howpublished = {\url{https://github.com/easylist/easylist/commit/0da40495282da4efe10a804a1307af5250650753}},
  note         = {Adds the filters \texttt{/get\_cookie\_id?cookie\_domain=} to \texttt{easyprivacy\_general.txt}; \texttt{||tagging.conde.digital\textasciicircum{}} to \texttt{easyprivacy\_thirdparty.txt}. Accessed: 2026-11-09}
}

@misc{easyprivacy2026-commit-663da24,
  author       = {{ryanbr}},
  title        = {{EasyPrivacy Commit 663da24}},
  year         = {2026},
  month        = nov,
  day          = {8},
  howpublished = {\url{https://github.com/easylist/easylist/commit/663da24801c60d1f8fbac9a7a0f468b6e2de2ed7}},
  note         = {Adds the filters \texttt{/v1/external/visitor|} and \texttt{://sgtm.} to \texttt{easyprivacy\_general.txt}; \texttt{||absmartly.io\textasciicircum{}\$third-party} to \texttt{easyprivacy\_trackingservers\_thirdparty.txt}. Accessed: 2026-11-09}
}

@misc{easyprivacy2026-commit-96f8851,
  author       = {{ryanbr}},
  title        = {{EasyPrivacy Commit 96f8851}},
  year         = {2026},
  month        = nov,
  day          = {8},
  howpublished = {\url{https://github.com/easylist/easylist/commit/96f8851acddc2e73b63581b50a7b9531f33ae6d5}},
  note         = {Adds the filter \texttt{/e?k=*\&v=\$image} to \texttt{easyprivacy\_general.txt}. Accessed: 2026-11-09}
}

@misc{easyprivacy2026-commit-42e7847,
  author       = {{ryanbr}},
  title        = {{EasyPrivacy Commit 42e7847}},
  year         = {2026},
  month        = nov,
  day          = {8},
  howpublished = {\url{https://github.com/easylist/easylist/commit/42e7847d833d1c0f0f470d7185d13bfd0c21b85c}},
  note         = {Adds the filter \texttt{/ccm/collect?} to \texttt{easyprivacy\_general.txt}. Accessed: 2026-11-09}
}

@misc{easyprivacy2026-commit-ec812ae,
  author       = {{ryanbr}},
  title        = {{EasyPrivacy Commit ec812ae}},
  year         = {2026},
  month        = nov,
  day          = {8},
  howpublished = {\url{https://github.com/easylist/easylist/commit/ec812ae813e87a920fb2f3dd2fb732b0f8c23539}},
  note         = {Adds the filter \texttt{://gtm.prod.} to \texttt{easyprivacy\_general.txt}. Accessed: 2026-11-09}
}

@misc{easyprivacy2026-commit-e2b53d3,
  author       = {{ryanbr}},
  title        = {{EasyPrivacy Commit e2b53d3}},
  year         = {2026},
  month        = nov,
  day          = {8},
  howpublished = {\url{https://github.com/easylist/easylist/commit/e2b53d3d7ac218be5d928aaf9b8f91d9cf8f3c00}},
  note         = {Adds the filters \texttt{\&gtm=*\&tag\_exp=}, \texttt{/a?id=*\&gtm=}, \texttt{/ddm/activity/src=}, and \texttt{/measurement/conversion?} to \texttt{easyprivacy\_general.txt}. Accessed: 2026-11-09}
}

@misc{easyprivacy2026-commit-d5d631f,
  author       = {{ryanbr}},
  title        = {{EasyPrivacy Commit d5d631f}},
  year         = {2026},
  month        = nov,
  day          = {8},
  howpublished = {\url{https://github.com/easylist/easylist/commit/d5d631f4cfb3c076bce028472ba8140563fc96a3}},
  note         = {Adds the filter \texttt{?id=GTM-*\&l=gtmDataLayer} to \texttt{easyprivacy\_general.txt}. Accessed: 2026-11-09}
}

@misc{easyprivacy2026-commit-1a11ea3,
  author       = {{ryanbr}},
  title        = {{EasyPrivacy Commit 1a11ea3}},
  year         = {2026},
  month        = nov,
  day          = {8},
  howpublished = {\url{https://github.com/easylist/easylist/commit/1a11ea35658779b96d7c73d4d2553ac147ec5ce4}},
  note         = {Adds the filter \texttt{://load.sgtm.} to \texttt{easyprivacy\_general.txt}. Accessed: 2026-11-09}
}

@article{jazlan2026tracking,
  title={Tracking Conversations: Measuring Content and Identity Exposure on AI Chatbots},
  author={Jazlan, Muhammad and Wang, Ethan and Vekaria, Yash and Shafiq, Zubair},
  journal={arXiv preprint arXiv:2604.27438},
  year={2026}
}

@misc{netflixCAPI,
	author = {},
	title = {{N}etflix {A}ds {S}uite {E}xpands {C}apabilities - {A}bout {N}etflix --- about.netflix.com},
	howpublished = {\url{https://about.netflix.com/en/news/netflix-ads-suite-expands-capabilities}},
	year = {n.d.},
	note = {[Accessed 28-03-2026]},
}

@misc{microsoftCAPI,
	author = {jonmeyers},
	title = {{C}onversions {A}{P}{I} ({C}{A}{P}{I}) {G}uide - {M}icrosoft {A}dvertising {A}{P}{I} --- learn.microsoft.com},
	howpublished = {\url{https://learn.microsoft.com/en-us/advertising/guides/uet-conversion-api-integration}},
	year = {n.d.},
	note = {[Accessed 28-03-2026]},
}

@misc{freshpaintGoogleAnalytics,
	author = {},
	title = {{G}oogle {A}nalytics 4 {S}erver-{S}ide {R}eference | {F}reshpaint --- documentation.freshpaint.io},
	howpublished = {\url{https://documentation.freshpaint.io/integrations/destinations/apps/google-analytics-4/google-analytics-4-server-side-reference}},
	year = {n.d.},
	note = {[Accessed 23-02-2026]},
}

@misc{wilson2024DMV,
	author = {Katherine Wilson, Class Members},
	title = {{K}atherine {W}ilson v. {G}oogle {L}{L}{C}},
	howpublished = {\url{https://cdn.arstechnica.net/wp-content/uploads/2024/05/Wilson-v-Google-Complaint-5-24-2024.pdf}},
	year = {},
	note = {[Accessed 21-02-2026]},
}

@misc{betterhelp2023FTC,
	author = {{Federal Trade Commission}},
	title = {In the Matter of {BetterHelp, Inc}},
	howpublished = {\url{https://www.ftc.gov/system/files/ftc\_gov/pdf/2023169betterhelpcomplaintfinal.pdf}},
	year = {2023},
	note = {[Accessed 22-02-2026]},
}

@misc{flo2021FTC,
	author = {{Federal Trade Commission}},
	title = {In the Matter of {Flo Health, Inc.}},
	howpublished = {\url{https://www.ftc.gov/system/files/documents/cases/192\_3133\_flo\_health\_complaint.pdf}},
	year = {2021},
	note = {[Accessed 22-02-2026]},
}

@misc{CDPnetworkDomain,
	author = {},
	title = {{C}hrome {D}ev{T}ools {P}rotocol --- chromedevtools.github.io},
	howpublished = {\url{https://chromedevtools.github.io/devtools-protocol/tot/Network/\#type-Initiator}},
	year = {n.d.},
	note = {[Accessed 21-02-2026]},
}

@misc{trancoList,
	author = {},
	title = {{A} research-oriented top sites ranking hardened against manipulation - {T}ranco --- tranco-list.eu},
	howpublished = {\url{https://tranco-list.eu/}},
	year = {n.d.},
	note = {[Accessed 17-02-2026]},
}

@article{dao2021cname,
  title={CNAME cloaking-based tracking on the web: Characterization, detection, and protection},
  author={Dao, Ha and Mazel, Johan and Fukuda, Kensuke},
  journal={IEEE Transactions on Network and Service Management},
  volume={18},
  number={3},
  pages={3873--3888},
  year={2021},
  publisher={IEEE}
}

@article{demir2022towards,
  title={Towards understanding first-party cookie tracking in the field},
  author={Demir, Nurullah and Theis, Daniel and Urban, Tobias and Pohlmann, Norbert},
  journal={arXiv preprint arXiv:2202.01498},
  year={2022}
}

@inproceedings{chen2021cookie,
  title={Cookie swap party: Abusing first-party cookies for web tracking},
  author={Chen, Quan and Ilia, Panagiotis and Polychronakis, Michalis and Kapravelos, Alexandros},
  booktitle={Proceedings of the Web Conference 2021},
  pages={2117--2129},
  year={2021}
}

@misc{stapeLoadContainer,
	author = {},
	title = {{L}oad the web {G}{T}{M} container from your subdomain and modify the request path --- stape.io},
	howpublished = {\url{https://stape.io/solutions/custom-gtm-loader}},
	year = {n.d.},
	note = {[Accessed 04-02-2026]},
}

@article{vekaria2025sok,
  title={SoK: Advances and Open Problems in Web Tracking},
  author={Vekaria, Yash and Beugin, Yohan and Munir, Shaoor and Acar, Gunes and Bielova, Nataliia and Englehardt, Steven and Iqbal, Umar and Kapravelos, Alexandros and Laperdrix, Pierre and Nikiforakis, Nick and others},
  journal={arXiv preprint arXiv:2506.14057},
  year={2025}
}

@inproceedings{fouad2024devil,
  title={The Devil is in the Details: Detection, Measurement and Lawfulness of Server-Side Tracking on the Web},
  author={Fouad, Imane and Santos, Cristiana and Laperdrix, Pierre},
  booktitle={24th Privacy Enhancing Technologies Symposium (PETS 2024)},
  volume={2024},
  number={4},
  year={2024}
}

@article{ghani2026pixelconfig,
  title={PixelConfig: Longitudinal Measurement and Reverse-Engineering of Meta Pixel Configurations},
  author={Ghani, Abdullah and Vekaria, Yash and Shafiq, Zubair},
  journal={arXiv preprint arXiv:2603.09380},
  year={2026}
}

@inproceedings{munir2023cookiegraph,
  title={Cookiegraph: Understanding and detecting first-party tracking cookies},
  author={Munir, Shaoor and Siby, Sandra and Iqbal, Umar and Englehardt, Steven and Shafiq, Zubair and Troncoso, Carmela},
  booktitle={Proceedings of the 2023 ACM SIGSAC Conference on Computer and Communications Security},
  pages={3490--3504},
  year={2023}
}

@inproceedings{nikkhah2025cookieguard,
  title={CookieGuard: Characterizing and Isolating the First-Party Cookie Jar},
  author={Nikkhah Bahrami, Pouneh and Fass, Aurore and Shafiq, Zubair},
  booktitle={Proceedings of the 2025 ACM Internet Measurement Conference},
  pages={645--661},
  year={2025}
}

@misc{stapeAvoidBlocking,
	author = {},
	title = {{H}ow to avoid {G}{A}4 and {G}{T}{M} blocking by ad blockers | {S}tape --- stape.io},
	howpublished = {\url{https://stape.io/blog/avoiding-google-tag-manager-ga4-blocking-by-adblockers}},
	year = {n.d.},
	note = {[Accessed 04-02-2026]},
}

@misc{Fullstorydevfullstorydatalayerobserver,
	author = {},
	title = {{G}it{H}ub - fullstorydev/fullstory-data-layer-observer: {O}bserve, transform, and send data layer content to {F}ull{S}tory --- github.com},
	howpublished = {\url{https://github.com/fullstorydev/fullstory-data-layer-observer}},
	year = {n.d.},
	note = {[Accessed 04-02-2026]},
}

@misc{matomoListFeatures,
	author = {},
	title = {{L}ist of {A}ll {F}eatures in {M}atomo {A}nalytics --- matomo.org},
	howpublished = {\url{https://matomo.org/features/}},
	year = {n.d.},
	note = {[Accessed 04-02-2026]},
}

@misc{docsAdblockRust,
	author = {},
	title = {adblock - {R}ust --- docs.rs},
	howpublished = {\url{https://docs.rs/adblock/latest/adblock/}},
	year = {n.d.},
	note = {[Accessed 03-02-2026]},
}

@misc{facebookIntegrateConversionsWithSGTM,
	author = {},
	title = {{I}ntegrate {C}onversions {A}{P}{I} with {S}erver-side {G}oogle {T}ag {M}anager in {E}vents {M}anager | {M}eta {B}usiness {H}elp {C}enter --- facebook.com},
	howpublished = {\url{https://www.facebook.com/business/help/702509907046774}},
	year = {n.d.},
	note = {[Accessed 03-02-2026]},
}

@misc{gaDuplicateEventsBlog2026,
	author = {},
	title = {{H}ow to fix duplicate events in {G}{A}4 - {O}ptimize {S}mart --- optimizesmart.com},
	howpublished = {\url{https://www.optimizesmart.com/how-to-fix-duplicate-events-in-ga4/}},
	year = {n.d.},
	note = {[Accessed 03-02-2026]},
}

@misc{caidaOrganizationsMappings,
	author = {},
	title = {{A}{S} to organizations mappings ({A}{S}2{O}rg) --- catalog.caida.org},
	howpublished = {\url{https://catalog.caida.org/dataset/as\_organizations}},
	year = {n.d.},
	note = {[Accessed 03-02-2026]},
}

@misc{teamcymruMappingService,
	author = {},
	title = {{I}{P} to {A}{S}{N} {M}apping {S}ervice | {T}eam {C}ymru --- team-cymru.com},
	howpublished = {\url{https://www.team-cymru.com/ip-asn-mapping}},
	year = {n.d.},
	note = {[Accessed 03-02-2026]},
}

@inproceedings{darkPatternsAfterGDPR2020,
    author = {Nouwens, Midas and Liccardi, Ilaria and Veale, Michael and Karger, David and Kagal, Lalana},
    title = {Dark Patterns after the GDPR: Scraping Consent Pop-ups and Demonstrating their Influence},
    year = {2020},
    isbn = {9781450367080},
    publisher = {Association for Computing Machinery},
    address = {New York, NY, USA},
    url = {https://doi.org/10.1145/3313831.3376321},
    doi = {10.1145/3313831.3376321},
    booktitle = {Proceedings of the 2020 CHI Conference on Human Factors in Computing Systems},
    pages = {1–13},
    numpages = {13},
    location = {Honolulu, HI, USA},
    series = {CHI '20}
}

@inproceedings{cookiescanner2023, series={ARES 2023},
   title={Cookiescanner: An Automated Tool for Detecting and Evaluating GDPR Consent Notices on Websites},
   url={http://dx.doi.org/10.1145/3600160.3605000},
   DOI={10.1145/3600160.3605000},
   booktitle={Proceedings of the 18th International Conference on Availability, Reliability and Security},
   publisher={ACM},
   author={Gundelach, Ralf and Herrmann, Dominik},
   year={2023},
   month=aug, pages={1–8},
   collection={ARES 2023} }

@inproceedings {cookieNoticeCompliance2024,
author = {Ahmed Bouhoula and Karel Kubicek and Amit Zac and Carlos Cotrini and David Basin},
title = {Automated {Large-Scale} Analysis of Cookie Notice Compliance},
booktitle = {33rd USENIX Security Symposium (USENIX Security 24)},
year = {2024},
isbn = {978-1-939133-44-1},
address = {Philadelphia, PA},
pages = {1723--1739},
url = {https://www.usenix.org/conference/usenixsecurity24/presentation/bouhoula},
publisher = {USENIX Association},
month = aug
}

@misc{googleGoogleAssistant,
	author = {},
	title = {Google Tag Assistant --- tagassistant.google.com},
	howpublished = {\url{https://tagassistant.google.com}},
	year = {n.d.},
	note = {[Accessed 31-01-2026]},
}

@misc{capiLaunchDateBlog,
	author = {Jessica Taylor},
	title = {{E}verything {Y}ou {N}eed {T}o {K}now {A}bout {F}acebook {C}onversions {A}{P}{I}},
	howpublished = {\url{https://portent.com/blog/paid-social/everything-you-need-to-know-about-facebook-conversions-api.html}},
	year = {n.d.},
	note = {[Accessed 30-01-2026]},
}

@misc{simoahavaServersideTaggingLaunchBlog,
	author = {Simo Ahava},
	title = {{S}erver-side {T}agging {I}n {G}oogle {T}ag {M}anager --- simoahava.com},
	howpublished = {\url{https://www.simoahava.com/analytics/server-side-tagging-google-tag-manager/}},
	year = {n.d.},
	note = {[Accessed 31-01-2026]},
}

@misc{iabtechlabTechTrusted,
	author = {},
	title = {{T}ech {L}ab {T}rusted {S}erver --- iabtechlab.com},
	howpublished = {\url{https://iabtechlab.com/tech-lab-trusted-server/}},
	year = {n.d.},
	note = {[Accessed 31-01-2026]},
}

@article{mertens2026detecting,
  title={Detecting and Measuring Client-and Server-Side Google Tag Manager and its Tags in 80K Websites},
  author={Mertens, Gilles and Bielova, Nataliia and Roca, Vincent and Bouhoula, Ahmed and Akassab, Marouanne},
  year={2026}
}

@inproceedings{mertens2025you,
  title={You Can't Trust Your Tag Neither: Privacy Leaks and Potential Legal Violations within the Google Tag Manager},
  author={Mertens, Gilles and Bielova, Nataliia and Roca, Vincent and Santos, Cristiana},
  booktitle={EuroS\&P 2025-10th IEEE European Symposium on Security and Privacy},
  year={2025}
}

@article{moti2025bitter,
  title={The Bitter Pill: Tracking and Remarketing on EU Pharmacy Websites},
  author={Moti, Zahra and Frings, Kimberley and Utz, Christine and Borgesius, Frederik Zuiderveen and Acar, Gunes},
  journal={Data Privacy Management. https://gunesacar. net/assets/bitter-pill-pharmacy-privacy-dpm-25. pdf},
  year={2025}
}

@misc{googleIntroductionTagging,
	author = {},
	title = {{I}ntroduction to tagging and the {G}oogle tag | {T}ag {P}latform | {G}oogle for {D}evelopers --- developers.google.com},
	howpublished = {\url{https://developers.google.com/tag-platform/devguides}},
	year = {n.d.},
	note = {[Accessed 30-01-2026]},
}

@misc{cookiepedia,
	author = {},
	title = {{A}ll {Y}ou {N}eed to {K}now {A}bout {C}ookies | {C}ookiepedia --- cookiepedia.co.uk},
	howpublished = {\url{https://cookiepedia.co.uk}},
	year = {n.d.},
	note = {[Accessed 26-01-2026]},
}

@inbook{WebAlmanac.2025.ThirdParties,
    author = "Jazlan, Muhammad and Aziz, Muhammad Abu Bakar and Pollard, Barry",
    title = "Third Parties",
    booktitle = "The 2025 Web Almanac",
    chapter = 3,
    publisher = "HTTP Archive",
    year = "2025",
    language = "English",
    doi = "10.5281/zenodo.18246420",
    url = "https://almanac.httparchive.org/en/2025/third-parties"
}

@misc{playwright,
	author = {},
	title = {{F}ast and reliable end-to-end testing for modern web apps | {P}laywright --- playwright.dev},
	howpublished = {\url{https://playwright.dev}},
	year = {n.d.},
	note = {[Accessed 01-01-2026]},
}

@misc{consent-o-matic,
	author = {},
	title = {{G}it{H}ub - cavi-au/{C}onsent-{O}-{M}atic: {B}rowser extension that automatically fills out cookie popups based on your preferences --- github.com},
	howpublished = {\url{https://github.com/cavi-au/Consent-O-Matic}},
	year = {n.d.},
	note = {[Accessed 01-01-2026]},
}

@misc{google_sgtm_why_when,
  author = {{Google for Developers}},
  title = {Why and when to use server-side tagging?},
  year = {2024},
  howpublished = {\url{https://developers.google.com/tag-platform/learn/sst-fundamentals/3-why-and-when-sst}},
  note = {Accessed: 2025-12-17}
}

@misc{reddit_capi,
  author = {{Reddit for Business}},
  title = {Conversions API},
  year = {2024},
  howpublished = {\url{https://business.reddithelp.com/s/article/Conversions-API}},
  note = {Accessed: 2025-12-17}
}

@misc{snap_capi,
  author = {{Snap for Developers}},
  title = {Conversions API},
  year = {2024},
  howpublished = {\url{https://developers.snap.com/api/marketing-api/Conversions-API/Introduction}},
  note = {Accessed: 2025-12-17}
}

@misc{tiktok_events_api,
  author = {{TikTok for Business}},
  title = {About Events API},
  year = {2025},
  howpublished = {\url{https://ads.tiktok.com/help/article/events-api}},
  note = {Accessed: 2025-12-17}
}

@misc{meta_capi_about,
  author = {{Meta Business Help Center}},
  title = {About Conversions API},
  year = {2025},
  howpublished = {\url{https://www.facebook.com/business/help/2041148702652965}},
  note = {Accessed: 2025-12-17}
}

@inproceedings{elfraihi2024serverside,
  author = {Asmaa El fraihi and Nardjes Amieur and Walter Rudametkin and Oana Goga},
  title = {Client-side and Server-side Tracking on Meta: Effectiveness and Accuracy},
  booktitle = {Proceedings on Privacy Enhancing Technologies},
  year = {2024},
  volume = {2024},
  number = {3},
  pages = {431--445},
  doi = {10.56553/popets-2024-0086}
}

@inproceedings{bekos2025piixel,
  title={PIIxel Leaks: Passive Identification of Personally Identifiable Information Leakage through Meta Pixel},
  author={Bekos, Paschalis and Papadopoulos, Panagiotis and Kourtellis, Nicolas and Polychronakis, Michalis},
  booktitle={Proceedings of the 2025 ACM SIGSAC Conference on Computer and Communications Security},
  pages={4229--4243},
  year={2025}
}

@inproceedings{bekos2023hitchhiker,
  title={The Hitchhiker’s guide to facebook web tracking with invisible pixels and click IDs},
  author={Bekos, Paschalis and Papadopoulos, Panagiotis and Markatos, Evangelos P and Kourtellis, Nicolas},
  booktitle={Proceedings of the ACM Web Conference 2023},
  pages={2132--2143},
  year={2023}
}

@article{kieserman2025tracker,
  title={Tracker Installations Are Not Created Equal: Understanding Tracker Configuration of Form Data Collection},
  author={Kieserman, Julia B and Andreou, Athanasios and Geeng, Chris and Lauinger, Tobias and McCoy, Damon},
  journal={arXiv preprint arXiv:2506.16891},
  year={2025}
}

@techreport{iab2024revenue,
  author = {{PwC and IAB}},
  title = {IAB Internet Advertising Revenue Report: Full Year 2024 Results},
  institution = {Interactive Advertising Bureau},
  year = {2024},
  url = {https://www.iab.com/wp-content/uploads/2025/04/IAB_PwC-Internet-Ad-Revenue-Report-Full-Year-2024.pdf}
}

@misc{fouad2020missedfilterlistsdetecting,
      title={Missed by Filter Lists: Detecting Unknown Third-Party Trackers with Invisible Pixels}, 
      author={Imane Fouad and Nataliia Bielova and Arnaud Legout and Natasa Sarafijanovic-Djukic},
      year={2020},
      eprint={1812.01514},
      archivePrefix={arXiv},
      primaryClass={cs.CR},
      url={https://arxiv.org/abs/1812.01514}, 
}

@misc{AboutAdBlockPlus,
  author       = {},
  title        = {{A}bout {A}dblock {P}lus --- adblockplus.org},
  howpublished = {\url{https://adblockplus.org/en/about}},
  year         = {},
  note         = {[Accessed 21-01-2025]}
}

@article{Iqbal2018adgraph,
  author     = {Umar Iqbal and Zubair Shafiq and Peter Snyder and Shitong Zhu and Zhiyun Qian and Benjamin Livshits},
  bibsource  = {dblp computer science bibliography, https://dblp.org},
  eprint     = {1805.09155},
  eprinttype = {arXiv},
  journal    = {CoRR},
  title      = {AdGraph: {A} Machine Learning Approach to Automatic and Effective Adblocking},
  url        = {http://arxiv.org/abs/1805.09155},
  volume     = {abs/1805.09155},
  year       = {2018}
}

@misc{BraveLists,
  author       = {Brave},
  title        = {{G}it{H}ub - brave/adblock-lists: {M}aintains adblock lists that {B}rave uses --- github.com},
  howpublished = {\url{https://github.com/brave/adblock-lists/}},
  year         = {},
  note         = {[Accessed 21-01-2025]}
}

@misc{BraveShields,
  author       = {},
  title        = {{B}rave {S}hields - {B}locking {A}ds, {T}rackers \& more | {B}rave - brave.com},
  howpublished = {\url{https://brave.com/shields/}},
  year         = {},
  note         = {[Accessed 21-01-2025]}
}

@article{dimova2021cname,
  title={The cname of the game: Large-scale analysis of dns-based tracking evasion},
  author={Dimova, Yana and Acar, Gunes and Olejnik, Lukasz and Joosen, Wouter and Van Goethem, Tom},
  journal={arXiv preprint arXiv:2102.09301},
  year={2021}
}

@misc{easylist,
  author       = {},
  title        = {EasyList},
  howpublished = {\url{https://easylist.to/easylist/easylist.txt}},
  year         = {2025},
  note         = {[Accessed 21-01-2025]}
}

@misc{easyprivacy,
  author       = {},
  title        = {EasyPrivacy},
  howpublished = {\url{https://easylist.to/easylist/easyprivacy.txt}},
  year         = {2025},
  note         = {[Accessed 21-01-2025]}
}

@misc{mozillaEnhancedTracking,
  author       = {},
  title        = { {E}nhanced {T}racking {P}rotection in {F}irefox for desktop | {F}irefox {H}elp  --- support.mozilla.org},
  howpublished = {\url{https://support.mozilla.org/en-US/kb/enhanced-tracking-protection-firefox-desktop}},
  year         = {},
  note         = {[Accessed 21-01-2025]}
}

@inproceedings{sjosten2020filter,
  title={Filter list generation for underserved regions},
  author={Sj{\"o}sten, Alexander and Snyder, Peter and Pastor, Antonio and Papadopoulos, Panagiotis and Livshits, Benjamin},
  booktitle={Proceedings of The Web Conference 2020},
  pages={1682--1692},
  year={2020}
}

@inproceedings{munir2024purl,
  title={PURL: Safe and Effective Sanitization of Link Decoration},
  author={Munir, Shaoor and Lee, Patrick and Iqbal, Umar and Shafiq, Zubair and Siby, Sandra},
  booktitle={33rd USENIX Security Symposium (USENIX Security 24)},
  pages={4103--4120},
  year={2024}
}

@misc{SafariPrivacyPage,
  author       = {Apple},
  title        = {{S}afari {P}rivacy {W}ebpage},
  howpublished = {\url{https://www.apple.com/safari/privacy}},
  year         = {},
  note         = {[Accessed 21-01-2025]}
}

@misc{uBlockOrigin,
  author       = {uBlock},
  title        = {{G}it{H}ub - gorhill/u{B}lock: u{B}lock {O}rigin - {A}n efficient blocker for {C}hromium and {F}irefox. {F}ast and lean. --- github.com},
  howpublished = {\url{https://github.com/gorhill/uBlock}},
  year         = {},
  note         = {[Accessed 21-01-2025]}
}

@inproceedings{siby2022webgraph,
  title={WebGraph: Capturing advertising and tracking information flows for robust blocking},
  author={Siby, Sandra and Iqbal, Umar and Englehardt, Steven and Shafiq, Zubair and Troncoso, Carmela},
  booktitle={31st USENIX security symposium (USENIX Security 22)},
  pages={2875--2892},
  year={2022}
}

\appendices

\renewcommand{\topfraction}{0.9}
\renewcommand{\bottomfraction}{0.8}
\renewcommand{\textfraction}{0.07}
\renewcommand{\floatpagefraction}{0.8}
\setcounter{topnumber}{3}
\setcounter{bottomnumber}{2}
\setcounter{totalnumber}{4}

\section{Classification Errors Analysis}
\label{appendex:classification-errors}
We manually inspect the false positives and false negatives in Stage 2 to understand where \tool fails and why.

\subsection{False positives from shared identifiers}
We exclude requests to known third parties, such as Facebook, Yahoo, Mountain, and Taboola, since existing filterlists already block them and \tool is designed to supplement those lists.
Across the individual modalities other than cookies, we observe 14 false positives in total.
Most occur on websites that implement other Google services, such as PageAd, Google Conversion Linker (CCM), and Google Ads.
These services can share identifiers with Google Analytics for ads and conversion tracking (\eg tracking ID \texttt{tid}, event name \texttt{en}, GTM container ID \texttt{gtm}), which our value templates match.
Since requests from these services are already blocked by existing filterlists, such false positives have little impact in practice.

\subsection{False positives from GA present on a webpage}
A second class of false positives occurs when a website carries GA state in the browser but sends no GA traffic.
For the Window Variables classifier, we manually inspected all 5 affected webpages and found that these were misconfigured GTM or GA tags that initialize client-side state, such as window variables, but never send analytics requests.
The Cookies classifier is most affected, with 138 false-positive domains.
On a sample of these websites, we find GA cookies but no requests to GA endpoints, and Tag Assistant reports no GA or sGA implementation.
Therefore, these also appear to be residual cookies from misconfigured tags or tag managers.
We conservatively count all such cases as false positives, since they do not meet our Stage 2 criteria.
The false positives of the ensemble models (28 for the Combined Classifier and 4 for the Meta Classifier) are explained by errors from the underlying modalities.

\subsection{False Negatives}
For all modalities, we compile a list of false negatives and manually validate them by visiting each website with our \tool extension activated.
This initially revealed data collection failures rather than classification errors: our consent mechanism~\cite{consent-o-matic} did not interact with the consent banner correctly, so GA was never triggered, or our crawler was detected as a bot and we collected no data for classification.
This is a well documented issue with automated crawling~\cite{darkPatternsAfterGDPR2020, cookieNoticeCompliance2024, cookiescanner2023}.
We manually updated their labels before computing the results we report.

We had 81 remaining false negatives across all modalities, primarily in the network request modality.
On \textit{themeisle.com}, for example, our cookie and window classifiers detect sGA, but requests to \textit{data.themeisle.com} are base64 encoded, so the network request classifier does not.
The domain-level network request classifier shares this blind spot, since such requests provide no request-level signal to aggregate.
The remaining false negatives show up in the ensemble classifiers, which inherit signals from the network request classifier.

\section{Extension Performance}
\label{subsec:evaluation:extension}

The main overhead with \tool is incurred by the Request-level classifier, which runs the feature extraction and classification process for each request.
This can potentially increase Page Load Time (PLT), degrading user experience.
Other models run periodically only once the \texttt{DOMContentLoaded} event fires, hence do not impact the PLT of a webpage.
To evaluate the impact of \tool on user experience, we measure PLT across three configurations: vanilla Chrome (baseline), uBlock Origin Lite (general adblocking), and our extension.
This establishes both the overhead of real-time sGA detection relative to no extension, and the performance trade-off relative to a well-known filterlist-based extension.

\begin{figure}[!htbp]
    \centering
    \includegraphics[width=0.48\linewidth]{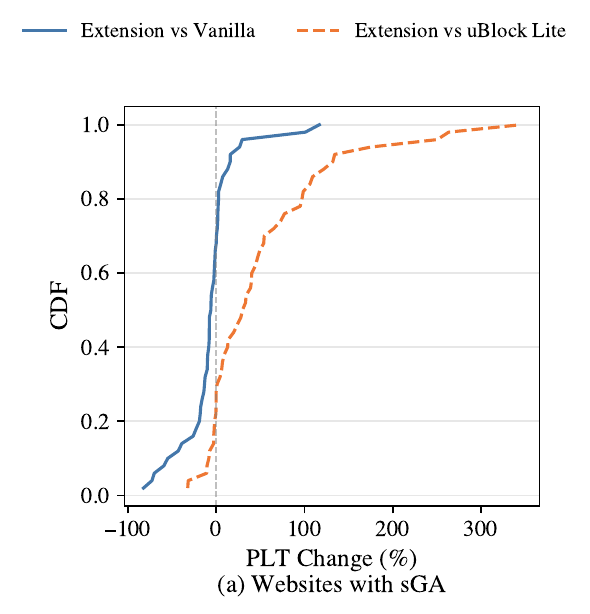}
    \hfill
    \includegraphics[width=0.48\linewidth]{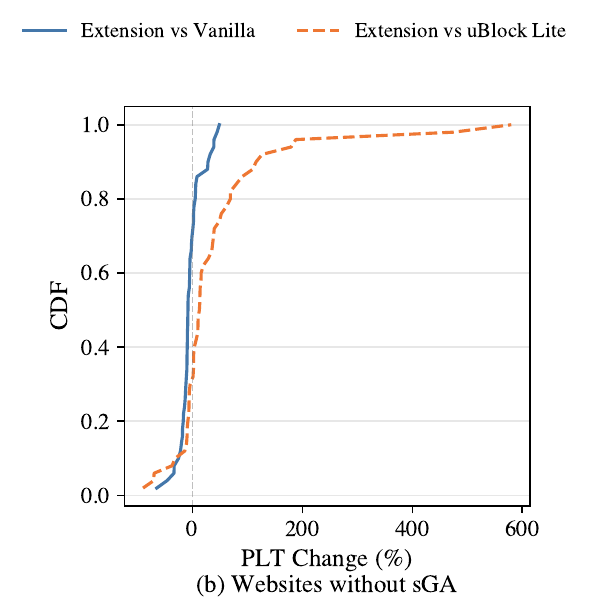}
    \caption{CDF of mean PLT change (vs.\ baseline) over 10 crawls across 50 websites with sGA (left) and 50 without sGA (right).}
    \label{fig:extension-performance}
\end{figure}

For our benchmarking, we visit 50 websites with sGA and 50 websites without sGA, performing one crawl per configuration.
To simulate realistic conditions, we limit the bandwidth to 10 Mbps with 100ms round-trip latency, following the approach established in prior work~\cite{Iqbal2018adgraph}.
We visit the landing page of each domain 10 times and record the PLT, defined as the interval between the DOM's \texttt{navigationStart} and \texttt{loadEventEnd} events, and report the average.
\Cref{fig:extension-performance} shows the difference in performance.

Across both sGA and non-sGA websites, \tool's mean PLT remains within approximately 10\% of vanilla Chrome on the vast majority of pages, well within typical run-to-run variation, as evidenced by the CDF concentration around 0\% PLT change in \Cref{fig:extension-performance}.
For fewer than 2\% of pages, the mean PLT deviates by more than 100\%, which can be attributed to network variability rather than systematic overhead.
uBlock Origin Lite, in contrast, improves PLT on roughly 80\% of websites; this is expected, since it blocks third-party content (\eg advertisements, analytics), reducing the resources the page must load before \texttt{loadEventEnd} fires.
These results demonstrate that \tool's real-time detection pipeline adds no observable overhead to user experience.
You can install the \tool extension from our artifacts at \url{https://github.com/jazlan01/sst-guard}

\section{Obfuscated URLs}
\Cref{tab:ssga-urls} shows examples of obfuscated sGA requests and their \texttt{base64} decoded versions.

\begin{table}[!htbp]
\centering
\caption{Examples of \texttt{base64} encoded sGA requests detected by \tool. All three encode the GA4 endpoint prefix \texttt{/g/collect?v=2\&tid=G-}.}
\label{tab:ssga-urls}
\footnotesize
\renewcommand{\arraystretch}{1.15}
\begin{tabular*}{\columnwidth}{@{}l@{\hspace{10pt}}l@{\extracolsep{\fill}}r@{}}
\toprule
\multicolumn{2}{@{}l}{\textbf{sGA endpoint}} & \textbf{DNS record} \\
\midrule
\multicolumn{2}{@{}l}{\texttt{data.themeisle.com}} & CNAME: \textit{usa.stape.io} \\
\quad Encoded & \multicolumn{2}{@{}l@{}}{\texttt{L2cvY29sbGVjdD92PTImdGlkPUct\ldots}} \\
\quad Decoded & \multicolumn{2}{@{}l@{}}{\texttt{/g/collect?v=2\&tid=G-D2Q8EGWPY5\ldots}} \\
\addlinespace[4pt]
\multicolumn{2}{@{}l}{\texttt{gtm.shapeways.com}} & A/AAAA \\
\quad Encoded & \multicolumn{2}{@{}l@{}}{\texttt{L2cvY29sbGVjdD92PTImdGlkPUct\ldots}} \\
\quad Decoded & \multicolumn{2}{@{}l@{}}{\texttt{/g/collect?v=2\&tid=G-MXZEPNTKS8\ldots}} \\
\addlinespace[4pt]
\multicolumn{2}{@{}l}{\texttt{assets.comparitech.com}} & CNAME: \textit{usc.stape.io} \\
\quad Encoded & \multicolumn{2}{@{}l@{}}{\texttt{L2cvY29sbGVjdD92PTImdGlkPUct\ldots}} \\
\quad Decoded & \multicolumn{2}{@{}l@{}}{\texttt{/g/collect?v=2\&tid=G-594Q6WX0ED\ldots}} \\
\bottomrule
\end{tabular*}
\end{table}
\newpage
\section{Network Level Analysis}
\begin{figure}[t]
    \centering
    \includegraphics[width=\linewidth]{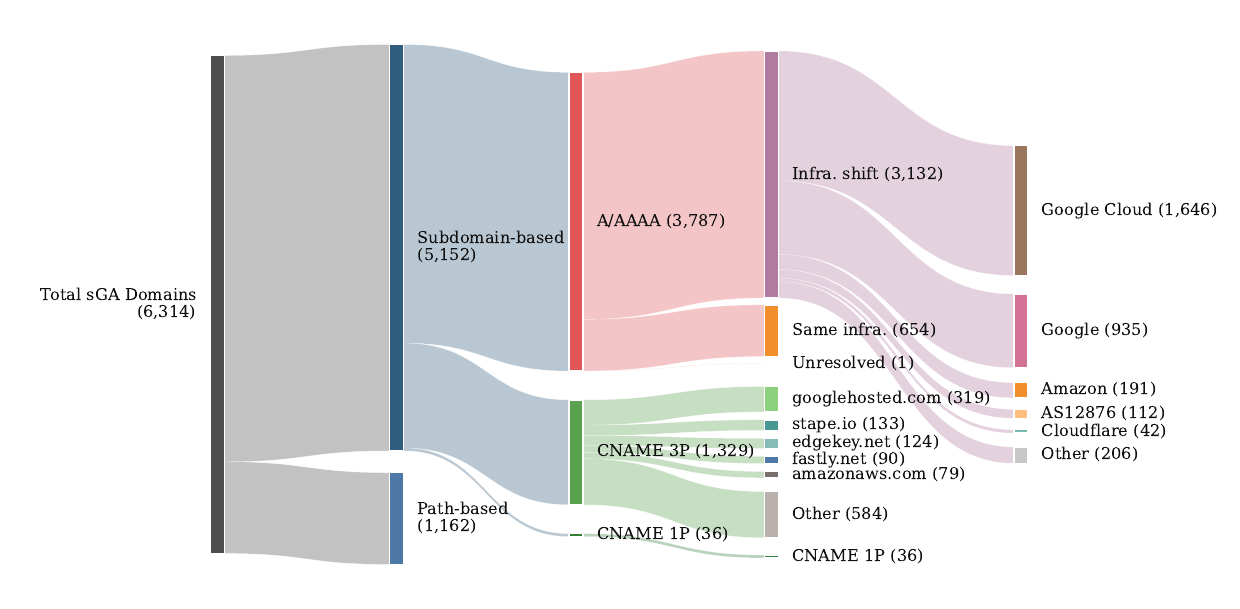}
    \caption{A Sankey diagram showing the findings of the network level analysis in \Cref{sec:measurement:network-level-analysis}.}
    \label{fig:sankey}
\end{figure}

\Cref{fig:sankey} summarizes the findings of our network level analysis in \Cref{sec:measurement:network-level-analysis}.

\section{Regexes}
\begin{table*}[p]
    \centering
    \caption{Complete Value-Template Regex Patterns by Modality}
    \label{tab:regexes}
    \setlength{\extrarowheight}{1pt}
    \begin{tabularx}{\linewidth}{@{}p{0.7cm}|p{3cm}>{\raggedright\arraybackslash\footnotesize}X@{}}
    \toprule
    & \textbf{Pattern Name} & \textbf{Regex Pattern} \\
    \midrule
    \multirow{5}{*}{\rotatebox{90}{\textit{Cookies}}}
      & standard\_ga     & \textasciicircum GA1\textbackslash{}.[123](-2)?\textbackslash{}.[0-9]\{6,10\}\textbackslash{}.17[0-9]\{8,13\}\$ \\
      & double\_prefix   & \textasciicircum GA1\textbackslash{}.1\textbackslash{}.GA1\textbackslash{}.2\textbackslash{}.[0-9]\{9,10\}\textbackslash{}.17[0-9]\{11\}\$ \\
      & alphanumeric     & \textasciicircum GA1\textbackslash{}.2\textbackslash{}.[a-z]\{3\}\textbackslash{}.[A-Za-z0-9]\{11\}\$ \\
      & uuid\_format     & \textasciicircum GA1\textbackslash{}.1\textbackslash{}.[a-z0-9]\{8\}-([0-9a-z]\{4\}-)\{3\}[0-9a-z]\{12\}\$ \\
      & ga4\_session     & \textasciicircum GS2\textbackslash{}.1\textbackslash{}.s17[0-9]\{8\}(\textbackslash{}\$[a-z][0-9]+)+\$ \\
    \midrule
    \multirow{10}{*}{\rotatebox{90}{\textit{Window Variables}}}
      & dataLayer\_event & "event":\textbackslash{}s*"gtm\textbackslash{}.(dom|load|js|scrollDepth)"\allowbreak|"event":\textbackslash{}s*"coreWebVitals" \\
      & gaGlobal\_hid    & "hid":\textbackslash{}s*\textbackslash{}d+ \\
      & gaGlobal\_vid    & "vid":\textbackslash{}s*"\textbackslash{}d+\textbackslash{}.17[0-9]\{8\}" \\
      & from\_cookie     & "from\textbackslash{}\_cookie":\textbackslash{}s*(?:true|false) \\
      & chrome\_version  & "144\textbackslash{}.0\textbackslash{}.7559\textbackslash{}.97" \\
      & brand\_strings   & "(Chromium|Google Chrome|Not\textbackslash{}\_A Brand)" \\
      & architecture     & "arm" \\
      & bitness          & "64" \\
      & platform\_version & "26\textbackslash{}.2\textbackslash{}.0" \\
      & container\_id    & "G-[A-Z0-9]\{5,10\}" \\
    \midrule
    \multirow{23}{*}{\rotatebox{90}{\textit{Query Parameters}}}
      & cid              & (?:GA\textbackslash{}d+\textbackslash{}.\textbackslash{}d+\textbackslash{}.)?\allowbreak\textbackslash{}d\{8,10\}\textbackslash{}.17\textbackslash{}d\{8,11\} \\
      & tid              & G-[A-Z0-9]\{10\} \\
      & dl               & https:\textbackslash{}/\textbackslash{}/[\textasciicircum{}\textbackslash{}s\&\#]+ \\
      & gtm              & 45[A-Za-z]\{1,2\}[0-9]\{1,2\}\allowbreak[a-z0-9A-Z]\{1,13\}(v8|v9)[A-Za-z0-9]+ \\
      & ul               & ([A-Za-z]\{2\}[-\textbackslash{}\_][A-Za-z]\{2\})\allowbreak|([A-Za-z]\{2\})\allowbreak|([Ee]nglish) \\
      & tag\_exp         & [0-9]\{9\}(\textasciitilde[0-9]\{9\})\{8,\} \\
      & gcd              & 13([a-zA-Z\textbackslash{}\_]\{1\}\textbackslash{}d\{1\})\{5\} \\
      & sid              & \textbackslash{}d\{10\} \\
      & \_p              & \textbackslash{}d\{13\} \\
      & pscdl            & (noapi|denied) \\
      & tfd              & \textbackslash{}d\{3,4\} \\
      & uaa              & x86 \\
      & uab              & 64 \\
      & uafvl            & \textasciicircum{}(?:(?:\textbackslash{}[%
          \{"brand":"Not\textbackslash{})A;Brand",\allowbreak"version":"\textbackslash{}d(?:\textbackslash{}.\textbackslash{}d)\{3\}"\},\allowbreak%
          \{"brand":"Chromium",\allowbreak"version":"\textbackslash{}d\{3\}(?:\textbackslash{}.\textbackslash{}d\textbackslash{}.\textbackslash{}d\{3\})\{2\}"\},\allowbreak%
          \{"brand":"Go\{2\}gle Chrome",\allowbreak"version":"\textbackslash{}d\{3\}(?:\textbackslash{}.\textbackslash{}d\textbackslash{}.\textbackslash{}d\{3\})\{2\}"\}%
          \textbackslash{}]\allowbreak%
          |Not\textbackslash{})A\%\textbackslash{}dB\{2\}rand\allowbreak%
          (?:\%\textbackslash{}dB\textbackslash{}d(?:\textbackslash{}.\textbackslash{}d)\{3\}\allowbreak%
          \%\textbackslash{}dC\{2\}hromium\allowbreak%
          \%\textbackslash{}dB\textbackslash{}d\{3\}(?:\textbackslash{}.\textbackslash{}d\textbackslash{}.\textbackslash{}d\{3\})\{2\}\allowbreak%
          \%\textbackslash{}dCGo\{2\}gle\%\textbackslash{}d\{2\}Chrome\%\textbackslash{}dB\allowbreak%
          |;\textbackslash{}d(?:\textbackslash{}.\textbackslash{}d)\{3\}%
          \textbackslash{}|Chromium;\textbackslash{}d\{3\}(?:\textbackslash{}.\textbackslash{}d\textbackslash{}.\textbackslash{}d\{3\})\{2\}\allowbreak%
          \textbackslash{}|Go\{2\}gle\%\textbackslash{}d\{2\}Chrome;)\allowbreak%
          \textbackslash{}d\{3\}(?:\textbackslash{}.\textbackslash{}d\textbackslash{}.\textbackslash{}d\{3\})\{2\})\$ \\
      & uap              & Linux \\
      & uapv             & 5\textbackslash{}.15\textbackslash{}.0 \\
      & en               & (page\textbackslash{}\_view\allowbreak|scroll\allowbreak|ad\textbackslash{}\_impression\allowbreak|user\textbackslash{}\_engagement\allowbreak|view\textbackslash{}\_item\textbackslash{}\_list\allowbreak|view\textbackslash{}\_item\allowbreak|scroll\textbackslash{}\_depth\allowbreak|view\textbackslash{}\_promotion\allowbreak|scroll\textbackslash{}\_75\allowbreak|time\textbackslash{}\_engaged\allowbreak|mp\textbackslash{}\_pageview\allowbreak|ddm\textbackslash{}\_standard\textbackslash{}\_event\allowbreak|click\allowbreak|Scroll Depth\allowbreak|page\textbackslash{}\_load\textbackslash{}\_time\allowbreak|scroll\textbackslash{}\_25\allowbreak|ads\textbackslash{}\_impression\allowbreak|scroll\textbackslash{}\_50\allowbreak|scroll\textbackslash{}\_tracking\allowbreak|proctor\allowbreak|Newsfeed\textbackslash{}\_show\allowbreak|Playbook Fired\allowbreak|scroll\textbackslash{}\_90\allowbreak|page\textbackslash{}\_scroll) \\
      & \_gid            & \textbackslash{}d\{8,10\}\textbackslash{}.\textbackslash{}d\{9,10\} \\
      & \_u              & ([A-Za-z]\{17\}|[A-Za-z]\{14\}|[A-Za-z]\{9\}|[A-Za-z]\{16\})\textasciitilde \\
      & \_eu             & ([A-Za-z]\{2,3\}[A-Z]\{4\})|([A-Z]\{2,4\})|([A-Za-z]\{2,3\}) \\
      & gcs              & G[0-3-]\{3\} \\
      & tcfd             & [0-6]\{2\}[0-9a-zA-Z]\{2,3\}[a-z]?\$ \\
      & ep.user\_agent   & \textasciicircum([Mm]ozilla\textbackslash{}/\textbackslash{}d+\textbackslash{}.\textbackslash{}d+\textbackslash{}s+\textbackslash{}([\textasciicircum)]+\textbackslash{})\textbackslash{}s+.+)\$ \\
    \bottomrule
    \end{tabularx}
\end{table*}

\Cref{tab:regexes} shows the regex patterns we use for \tool.
The templates for \texttt{chrome\_version}, \texttt{brand\_strings}, \texttt{architecture}, \texttt{bitness}, and \texttt{platform\_version} depend on the user's system, so our extension instantiates them at runtime from the browser.
The values in the table are shown as an example from our crawls.

\end{document}